\documentstyle[12pt]{article}

\makeatletter
\newcounter{subfigure}[figure]          
\def\thesubfigure{(\alph{subfigure})}   
\def\@thesubfigure{\thesubfigure\space} 
\def\p@subfigure{\thefigure}            
\let\ext@subfigure\ext@figure           
\newcounter{lofdepth}                   
\def\l@subfigure{
  \@dottedxxxline{\ext@subfigure}{2}{3.9em}{2.3em}}
\newcounter{subtable}[table]            
\def\thesubtable{(\alph{subtable})}     
\def\@thesubtable{\thesubtable\space}   
\def\p@subtable{\thetable}              

\let\ext@subtable\ext@table             
\newcounter{lotdepth}                   
\def\l@subtable{
  \@dottedxxxline{\ext@subtable}{2}{3.9em}{2.3em}}
\def\subcapsize{\footnotesize} 
\def\subfigtopskip{10pt}       
\def\subfigbottomskip{10pt}    
\def\subfigcapskip{10pt}       
\def\subfigcapmargin{10pt}     
\def\@subfigcaptionlist{}      
\def\subfigure{%
  \bgroup                     
    \advance\c@figure\@ne     
    \refstepcounter{subfigure}
    \leavevmode               
    \@ifnextchar [%
      {\@subfloat{subfigure}}%
      {\@subfloat{subfigure}[\@empty]}}
\def\subtable{%
  \bgroup                     
    \advance\c@table\@ne      
    \refstepcounter{subtable}
    \leavevmode               
    \@ifnextchar [%
      {\@subfloat{subtable}}%
      {\@subfloat{subtable}[\@empty]}}
\def\@subfloat#1[#2]#3{%
    \setbox\@tempboxa \hbox{#3}%
    \@tempdima=\wd\@tempboxa
    \vtop{%
      \vbox{%
        \vskip\subfigtopskip
        \box\@tempboxa}%
      \ifx \@empty#2\relax \else
        \@subcaption{#1}{#2}%
        \vskip\subfigcapskip
        \setbox\@tempboxa \hbox{\subcapsize\csname @the#1\endcsname #2}%
        \@tempdimb=-\subfigcapmargin
        \multiply\@tempdimb\tw@
        \advance\@tempdimb\@tempdima
        \hbox to\@tempdima{%
          \hfil
          \ifdim \wd\@tempboxa >\@tempdimb
            \parbox[t]{\@tempdimb}{\subcapsize\csname @the#1\endcsname #2}%
          \else
            \box\@tempboxa
          \fi
          \hfil}%
      \fi
      \vskip\subfigbottomskip}%
  \egroup}
\def\@subcaption#1#2{%
  \begingroup
    \def\protect{\string\string\string}%
    \xdef\@subfigcaptionlist{%
      \@subfigcaptionlist,%
      {\protect\numberline {\@currentlabel}%
       \noexpand{\ignorespaces #2}}}%
  \endgroup}
\def\@dottedxxxline#1#2#3#4#5#6{%
  \ifnum #2>\csname c@#1depth\endcsname \else
    \@dottedtocline{0}{#3}{#4}{#5}{#6}
  \fi}
\let\@@caption\@caption
\long\def\@caption#1[#2]#3{%
  \@@caption{#1}[{#2}]{#3}
  \@for \@tempa:=\@subfigcaptionlist \do {
    \ifx\@empty\@tempa\relax \else
      \addcontentsline
        {\csname ext@sub#1\endcsname}%
        {sub#1}%
        {\@tempa}%
    \fi}%
  \gdef\@subfigcaptionlist{}}
\newread\epsffilein    
\newif\ifepsffileok    
\newif\ifepsfbbfound   
\newif\ifepsfverbose   
\newdimen\epsfxsize    
\newdimen\epsfysize    
\newdimen\epsftsize    
\newdimen\epsfrsize    
\newdimen\epsftmp      
\newdimen\pspoints     
\newdimen\epsfhbpsize  
\newdimen\epsfvbpsize  
\def\epsfbox#1{\global\def\epsfllx{72}\global\def\epsflly{72}%
   \global\def\epsfurx{540}\global\def\epsfury{720}%
   \def\lbracket{[}\def\testit{#1}\ifx\testit\lbracket
   \let\next=\epsfgetlitbb\else\let\next=\epsfnormal\fi\next{#1}}%
\def\epsfgetlitbb#1#2 #3 #4 #5]#6{\epsfgrab #2 #3 #4 #5 .\\%
   \input{#6frag}}%
\def\epsfnormal#1{\epsfgetbb{#1}\input{#1frag}}%
\def\epsfgetbb#1{%
\openin\epsffilein=#1
\ifeof\epsffilein\errmessage{I couldn't open #1, will ignore it}\else
   {\epsffileoktrue \chardef\other=12
    \def\do##1{\catcode`##1=\other}\dospecials \catcode`\ =10
    \loop
       \read\epsffilein to \epsffileline
       \ifeof\epsffilein\epsffileokfalse\else
          \expandafter\epsfaux\epsffileline:. \\%
       \fi
   \ifepsffileok\repeat
   \ifepsfbbfound\else
    \ifepsfverbose\message{No bounding box comment in #1; using defaults}\fi\fi
   }\closein\epsffilein\fi}%
\def\epsfsetgraph#1{%
   \epsfrsize=\epsfury\pspoints
   \advance\epsfrsize by-\epsflly\pspoints
   \epsftsize=\epsfurx\pspoints
   \advance\epsftsize by-\epsfllx\pspoints
   \epsfxsize\epsfsize\epsftsize\epsfrsize
   \ifnum\epsfxsize=0 \ifnum\epsfysize=0
      \epsfxsize=\epsftsize \epsfysize=\epsfrsize
     \else\epsftmp=\epsftsize \divide\epsftmp\epsfrsize
       \epsfxsize=\epsfysize \multiply\epsfxsize\epsftmp
       \multiply\epsftmp\epsfrsize \advance\epsftsize-\epsftmp
       \epsftmp=\epsfysize
       \loop \advance\epsftsize\epsftsize \divide\epsftmp 2
       \ifnum\epsftmp>0
          \ifnum\epsftsize<\epsfrsize\else
             \advance\epsftsize-\epsfrsize \advance\epsfxsize\epsftmp \fi
       \repeat
     \fi
   \else\epsftmp=\epsfrsize \divide\epsftmp\epsftsize
     \epsfysize=\epsfxsize \multiply\epsfysize\epsftmp
     \multiply\epsftmp\epsftsize \advance\epsfrsize-\epsftmp
     \epsftmp=\epsfxsize
     \loop \advance\epsfrsize\epsfrsize \divide\epsftmp 2
     \ifnum\epsftmp>0
        \ifnum\epsfrsize<\epsftsize\else
           \advance\epsfrsize-\epsftsize \advance\epsfysize\epsftmp \fi
     \repeat
   \fi
   \expandafter\ifx\csname psfrag\endcsname\relax\else
       \epsftsize=\epsfurx\pspoints
       \advance\epsftsize by-\epsfllx\pspoints
       \epsfrsize=1\pspoints
       \ifnum\epsfxsize=0
            \epsfhbpsize=1\pspoints
       \else\epsftmp=\epsfrsize \divide\epsftmp\epsftsize
         \epsfhbpsize=\epsfxsize \multiply\epsfhbpsize\epsftmp
         \multiply\epsftmp\epsftsize \advance\epsfrsize-\epsftmp
         \epsftmp=\epsfxsize
         \loop \advance\epsfrsize\epsfrsize \divide\epsftmp 2
         \ifnum\epsftmp>0
            \ifnum\epsfrsize<\epsftsize\else
               \advance\epsfrsize-\epsftsize \advance\epsfhbpsize\epsftmp \fi
         \repeat
        \fi
        \epsfvbpsize=\epsfhbpsize
    \fi
   \ifepsfverbose\message{#1: width=\the\epsfxsize, height=\the\epsfysize}\fi
   \epsffileokfalse%
   \expandafter\ifx\csname psfrag\endcsname\relax\else%
      \openin\epsffilein=#1frag
      \ifeof\epsffilein\else%
          \closein\epsffilein%
          \epsffileoktrue%
   \fi\fi%
   \epsftmp=10\epsfxsize \divide\epsftmp\pspoints
   \vbox to\epsfysize{\vfil\hbox to\epsfxsize{%
      \ifepsffileok
          \def\PsFragSpecialArgs{PSfile=#1 llx=\epsfllx\space lly=\epsflly %
             \space urx=\epsfurx\space ury=\epsfury\space rwi=\number\epsftmp}%
          \input{#1frag}%
      \else
          \includegraphics{#1}%
      \fi\hfil}}%
\epsfxsize=0pt\epsfysize=0pt}%
{\catcode`\%=12 \global\let\epsfpercent=
\long\def\epsfaux#1#2:#3\\{\ifx#1\epsfpercent
   \def\testit{#2}\ifx\testit\epsfbblit
      \epsfgrab #3 . . . \\%
      \epsffileokfalse
      \global\epsfbbfoundtrue
   \fi\else\ifx#1\par\else\epsffileokfalse\fi\fi}%
\def\epsfgrab #1 #2 #3 #4 #5\\{%
   \global\def\epsfllx{#1}\ifx\epsfllx\empty
      \epsfgrab #2 #3 #4 #5 .\\\else
   \global\def\epsflly{#2}%
   \global\def\epsfurx{#3}\global\def\epsfury{#4}\fi}%
\def\epsfsize#1#2{\epsfxsize}

\typeout{Style Option REVMACS Version 1 as of 31 Mar 1994}
\topmargin 0pt
\advance \topmargin by -\headheight
\advance \topmargin by -\headsep
\textheight 8.9in
\oddsidemargin 0pt
\evensidemargin \oddsidemargin
\marginparwidth 0.5in
\textwidth 6.5in

\def\lesssim{\mathrel{\lower2.5pt\hbox{$\textstyle<$}\atop\raise2.5pt\hbox{$\textstyle\sim$}}}

\def\pacs#1{}

\def\@authoraddress{}  \def\@title{} \def\@date{} \def\@preprint{}
\def\and{\unskip, }
\def\preprint#1{%
\def\@preprint{\noindent\hfill\hbox{#1}\vskip 10pt}%
}
\def\title#1{\gdef\@title{{\large\bf\centering\ignorespaces#1\vskip2.5pt}}}
\def\author#1{\expandafter\def\expandafter\@authoraddress\expandafter
{\@authoraddress %
\vskip1.5pc %
{\dimen0=-\prevdepth \advance\dimen0 by23pt
\nointerlineskip \rm\centering
\vrule height\dimen0 width0pt\relax\ignorespaces#1\par
}%
}%
}
\def\address#1{\expandafter\def\expandafter\@authoraddress\expandafter
{\@authoraddress{\small\it\centering\ignorespaces#1\par}}}
\def\date#1{\gdef\@date{{\small\rm\centering(\ignorespaces#1\unskip)\par}}}
\def\maketitle{\par
\begingroup
\let\cite\@bylinecite
\def\thefootnote{\fnsymbol{footnote}}%
\if@twocolumn
\twocolumn[\@maketitle\vskip2pc]%
\else
\newpage
\global\@topnum\z@ %
\@maketitle
\fi
\thispagestyle{plain}\@thanks
\endgroup
\def\thefootnote{\arabic{footnote}}%
\setcounter{footnote}{0}%
\let\maketitle\relax \let\@maketitle\relax
\let\@thanks\relax \let\@authoraddress\relax \let\@title\relax
\let\@date\relax \let\thanks\relax
}
\def\@maketitle{%
\@preprint
\@title
\ifdim\prevdepth=-1000pt \prevdepth0pt\fi
\@authoraddress
\@date
}
\def\thesection       {\Roman{section}}
\def\p@section        {}
\def\thesubsection    {\Alph{subsection}}
\def\p@subsection     {\thesection\,}

\def\p@subsubsection  {\thesection\,\thesubsection\,}

\newif\if@mainhead
\let\reset@font\relax
\def\section{\@mainheadtrue
\@startsection {section}{1}{\z@}{0.8cm plus1ex minus
 .2ex}{0.5cm plus1ex minus.2ex}{\reset@font\small\bf\centering}}
\def\subsection{\@mainheadfalse
\@startsection{subsection}{2}{\z@}{0.8cm plus1ex minus
 .2ex}{0.5cm plus1ex minus.2ex}{\reset@font\small\bf\centering}}
\def\subsubsection{\@mainheadfalse
\@startsection{subsubsection}{3}{\z@}{.8cm plus1ex minus
 .2ex}{0.5cm plus1ex minus.2ex}{\reset@font\small\it\centering}}
\def\paragraph{\@mainheadfalse
\@startsection{paragraph}{4}{\parindent}{\z@}{-1em}{\reset@font
\normalsize\it}}
\def\subparagraph{\@mainheadfalse
\@startsection{subparagraph}{4}{\parindent}{3.25ex plus1ex minus
 .2ex}{-1em}{\reset@font\normalsize\bf}}
\def\baselinestretch{1.5}
\def\setstretch#1{\renewcommand{\baselinestretch}{#1}}
\@ifundefined{selectfont}
{
\def\@setsize#1#2#3#4{\@nomath#1
   \let\@currsize#1\baselineskip
   #2\baselineskip\baselinestretch\baselineskip
   \parskip\baselinestretch\parskip
   \setbox\strutbox\hbox{\vrule height.7\baselineskip
      depth.3\baselineskip width\z@}
   \normalbaselineskip\baselineskip#3#4}
}
{
\def\@setsize#1#2#3#4{\@nomath#1%
   \let\@currsize#1\parskip\baselinestretch\parskip
    \baselineskip\baselinestretch\baselineskip
              \size{#4}{#2}\selectfont}
}
\skip\footins 20pt plus4pt minus4pt
\def\@xfloat#1[#2]{\ifhmode \@bsphack\@floatpenalty -\@Mii\else
   \@floatpenalty-\@Miii\fi\def\@captype{#1}\ifinner
      \@parmoderr\@floatpenalty\z@
    \else\@next\@currbox\@freelist{\@tempcnta\csname ftype@#1\endcsname
       \multiply\@tempcnta\@xxxii\advance\@tempcnta\sixt@@n
       \@tfor \@tempa :=#2\do
                        {\if\@tempa h\advance\@tempcnta \@ne\fi
                         \if\@tempa t\advance\@tempcnta \tw@\fi
                         \if\@tempa b\advance\@tempcnta 4\relax\fi
                         \if\@tempa p\advance\@tempcnta 8\relax\fi
         }\global\count\@currbox\@tempcnta}\@fltovf\fi
    \global\setbox\@currbox\vbox\bgroup
    \def\baselinestretch{1}\small\normalsize
    \boxmaxdepth\z@
    \hsize\columnwidth \@parboxrestore}
\long\def\@footnotetext#1{\insert\footins{\def\baselinestretch{1}\footnotesize
    \interlinepenalty\interfootnotelinepenalty
    \splittopskip\footnotesep
    \splitmaxdepth \dp\strutbox \floatingpenalty \@MM
    \hsize\columnwidth \@parboxrestore
   \edef\@currentlabel{\csname p@footnote\endcsname\@thefnmark}\@makefntext
    {\rule{\z@}{\footnotesep}\ignorespaces
      #1\strut}}}
\def\singlespace{%
\vskip\parskip%
\vskip\baselineskip%
\def\baselinestretch{1}%
\@ifundefined{selectfont}{%
\ifx\@currsize\normalsize\@normalsize\else\@currsize\fi%
}{
\ifx\@currsize\normalsize\@normalsize\else\@currsize\fi\setnew@baselineskip%
}
\vskip-\parskip%
\vskip-\baselineskip%
}

\def\spacing#1{\par%
 \def\baselinestretch{#1}%
 \ifx\@currsize\normalsize\@normalsize\else\@currsize\fi}

\everydisplay{
   \abovedisplayskip \baselinestretch\abovedisplayskip%
   \belowdisplayskip \abovedisplayskip%
   \abovedisplayshortskip \baselinestretch\abovedisplayshortskip%
   \belowdisplayshortskip  \baselinestretch\belowdisplayshortskip}
\setstretch{1}

\newcount\@minsofar
\newcount\@min
\newcount\@cite@temp
\def\@citex[#1]#2{%
\if@filesw \immediate \write \@auxout {\string \citation {#2}}\fi
\@tempcntb\m@ne \let\@h@ld\relax \def\@citea{}%
\@min\m@ne%
\@cite{%
  \@for \@citeb:=#2\do {\@ifundefined {b@\@citeb}%
    {\@h@ld\@citea\@tempcntb\m@ne{\bf ?}%
    \@warning {Citation `\@citeb ' on page \thepage \space undefined}}%
{\@minsofar\z@ \@for \@scan@cites:=#2\do {%
  \@ifundefined{b@\@scan@cites}%
    {\@cite@temp\m@ne}
    {\@cite@temp\number\csname b@\@scan@cites \endcsname \relax}%
\ifnum\@cite@temp > \@min
    \ifnum\@minsofar = \z@
      \@minsofar\number\@cite@temp
      \edef\@scan@copy{\@scan@cites}\else
    \ifnum\@cite@temp < \@minsofar
      \@minsofar\number\@cite@temp
      \edef\@scan@copy{\@scan@cites}\fi\fi\fi}\@tempcnta\@min
  \ifnum\@minsofar > \z@ 
    \advance\@tempcnta\@ne
    \@min\@minsofar
    \ifnum\@tempcnta=\@minsofar 
      \ifx\@h@ld\relax
        \edef \@h@ld{\@citea\csname b@\@scan@copy\endcsname}%
      \else \edef\@h@ld{\ifmmode{-}\else--\fi\csname b@\@scan@copy\endcsname}%
      \fi
    \else \@h@ld\@citea\csname b@\@scan@copy\endcsname
          \let\@h@ld\relax
  \fi 
\fi}%
\def\@citea{,\penalty\@highpenalty\,}}\@h@ld}{#1}}

\def\captionfont{}   
\long\def\@makecaption#1#2{\footnotesize
   \vskip 10pt
   \setbox\@tempboxa\hbox{\captionfont#1: #2}
   \ifdim \wd\@tempboxa >\hsize   
       {\captionfont#1: #2}\par   
     \else                        
       \hbox to\hsize{\hfil\box\@tempboxa\hfil}
   \fi}
\makeatother

\topmargin 0pt
\advance \topmargin by -\headheight
\advance \topmargin by -\headsep
\textheight 8.9in
\oddsidemargin 0pt
\evensidemargin \oddsidemargin
\marginparwidth 0.5in
\textwidth 6.5in

\begin{document}

\def\Tf{T_f}
\def\Tg{T_g}
\newlength{\figurewidth}
\setlength{\figurewidth}{.95\textwidth}
\def\insfig#1#2{\centerline{%
    \epsfxsize=#1\figurewidth\epsfbox{#2}}}

\begin{titlepage}

\rightline{chem-ph/9411008}

\vspace{1cm}

\begin{center}
{\bf\large Funnels, Pathways and the Energy Landscape of Protein Folding:
  A Synthesis}

\vspace{2\baselineskip}

{Joseph D. Bryngelson, \\
 {\it Physical~Sciences~Laboratory, \\
 Division~of~Computer~Research~and~Technology, \\
 National Institutes of Health, Bethesda, MD 20892,} \\[\baselineskip]
 Jos\'e Nelson Onuchic, Nicholas D. Socci, \\
 {\it Department of Physics-0319\\
 University of California at San Diego\\
 La Jolla, California 92093-0319}\\[\baselineskip]
 Peter G. Wolynes \\
 {\it School of Chemical Sciences and Beckman Institute,\\
 University of Illinois, \\
 Urbana, Illinois 61801}\\[\baselineskip]
 {\sf In press: {\em Proteins}}}

\end{center}

\end{titlepage}

\vspace{2\baselineskip}

\centerline{\bf Abstract} \vspace{\baselineskip}
\centerline{\hbox{\vbox{\hsize=6in
      \noindent The understanding, and even the description of protein
      folding is impeded by the complexity of the process.  Much of
      this complexity can be described and understood by taking a
      statistical approach to the energetics of protein conformation,
      that is, to the energy landscape.  The statistical energy
      landscape approach explains when and why unique behaviors, such
      as specific folding pathways, occur in some proteins and more
      generally explains the distinction between folding processes
      common to all sequences and those peculiar to individual
      sequences.  This approach also gives new, quantitative insights
      into the interpretation of experiments and simulations of
      protein folding thermodynamics and kinetics.  Specifically, the
      picture provides simple explanations for folding as a two-state
      first-order phase transition, for the origin of metastable
      collapsed unfolded states and for the curved Arrhenius plots
      observed in both laboratory experiments and discrete lattice
      simulations.  The relation of these quantitative ideas to
      folding pathways, to uni-exponential {\em vs.} multi-exponential
      behavior in protein folding experiments and to the effect of
      mutations on folding is also discussed.  The success of energy
      landscape ideas in protein structure prediction is also
      described.  The use of the energy landscape approach for
      analyzing data is illustrated with a quantitative analysis of
      some recent simulations, and a qualitative analysis of
      experiments on the folding of three proteins.  The work unifies
      several previously proposed ideas concerning the mechanism
      protein folding and delimits the regions of validity of these
      ideas under different thermodynamic conditions.}}}


\section{Introduction}

The apparent complexity of folded protein structures and the
extraordinary diversity of conformational states of unfolded proteins
makes challenging even the description of protein folding in atomistic
terms.  Soon after Anfinsen's classic experiments on renaturation of
unfolded proteins~\cite{anfinsen_pnas}, Levinthal recognized the
conceptual difficulty of a molecule searching at random through the
cosmologically large number of unfolded configurations to find the
folded structure in a biologically relevant time~\cite{levinthal}.  To
resolve this ``paradox,'' he postulated the notion of a protein
folding pathway.  The search for such a pathway is often stated as the
motive for experimental protein folding studies. On the other hand,
the existence of multiple parallel paths to the folded state has been
occasionally invoked~\cite{jigsaw}.  Recently, a new approach to
thinking about protein folding and about these issues specifically has
emerged based on the statistical characterization of the energy
landscape of folding proteins~\cite{bw_pnas,bw_jpc,bw_biopolymers}.

This paper presents the basic ideas of the statistical energy
landscape view of protein folding and relates them to the older
languages of protein folding pathways.  The use of statistics to
describe protein physical chemistry is quite natural, even though each
protein has a specific sequence, structure and function essential to
its biological activity.  The huge number of conformational states
immediately both allows and requires a statistical characterization.
In addition folding is a general behavior common to a large ensemble
of biological molecules.  Many different sequences fold to essentially
the same structure as witnessed by the extreme dissimilarities in
sequence which may be found in families of proteins such as
lysozyme~\cite{swiss_prot}. Thus for any given observed protein
tertiary structure, there is a statistical ensemble of biological
molecules which fold to it.  Many studies suggest that the dynamics of
many parts of the folding process are common to all of the sequences
of a given overall structure, while others are peculiar to individual
sequences. {\it Distinguishing folding processes common to all
  sequences from those peculiar to individual sequences is a major
  goal of physical theories of protein folding.} The statistical
energy landscape analysis will show which features are common and
which are specific taxonomic aspects of protein folding.

Depending on the statistical characteristics of the energy landscape,
either a unique folding pathway or multiple pathways may emerge.  A
biological relevance of the distinction between the two pictures is
that mutations can more dramatically affect the dynamics through
unique pathways than through multiple pathways.

The organization of this paper is as follows: in the next section we
describe the energy landscape of protein folding, discuss the
properties of smooth and rough energy landscapes, and indicate that it
appears that protein folding occurs on an energy landscape that is
intermediate between most smooth and most rough.  In Section Three, we
describe a simple protein folding model that interpolates between
these two limits and exhibits both the smooth and the rough energy
landscape properties that are present in folding proteins.  The
equilibrium thermodynamic properties of this model are also discussed
in this section.  Section Four starts with a short survey of the
differences between the kinetics of complex chemical processes, such
as protein folding, and the kinetics of the simple chemical processes
whose understanding forms the basis of the most commonly used reaction
rate theories.  We review how these common theories should be modified
to cope with the complexity of a process like protein folding.  Then
we present the necessary modifications of kinetics and apply them to
the simple protein folding model of Section Three.  Each scenario has
its own characteristic behavior.  The folding scenario observed in any
given experiment depends on the specific sequence and the refolding
conditions.  Section Five shows how the scenarios presented in Section
Four can be understood in terms of the phase diagram for protein
folding.  This phase diagram is also discussed in detail.  In the next
section we show how the energy landscape ideas can be used to analyze
data by presenting a rough but quantitative analysis of some computer
simulation data.  In the following section, Section Seven, we give a
flavor of the issues in energy landscape analysis of experimental data
through an examination of some previously published experimental
results.  We also present a tentative assignment of the folding
scenarios observed in these experiments.  The concluding section then
summarizes the results, and discusses the significance of the energy
landscape for understanding protein folding, for protein structure
prediction and for protein engineering.

\section{Smoothness, Roughness and the Topography of Energy
  Landscapes}

Protein folding is a complex process, typically occurring at a
constant pressure and temperature, involving important changes in the
structure of both the chain and the
solvent~\cite{ghelis_yon,pain_book}. The natural thermodynamic
potential for describing processes at constant pressure and
temperature is the Gibbs free energy~\cite{pauling,lewis_randall}, so
we will use an effective free energy that is a function of the
configuration of the protein to describe the protein-solvent system.
Notice that this description implicitly averages over the solvent
coordinates.  This averaging means that the forces that arise from
this potential function are temperature dependent.  To make these
considerations more concrete, consider the forces on two apolar groups
immersed in water.  The apolar-group-solvent-system has a lower free
energy if the two apolar groups are close to one another, so the
solvent-averaged free energy, mentioned above, has a minimum when the
two groups are close and becomes larger when the groups are further
apart~\cite{franks}.  The change in the solvent-averaged free energy
as a function of distance between the groups causes the groups to
attract one another.  This attraction is the hydrophobic force.  Since
the free energy of the apolar-group-solvent-system changes as the
temperature changes, likewise the solvent-averaged free energy and the
hydrophobic force also change~\cite{franks,leikin}.

The need to consider the form of the free energy as a function of
protein conformation, which we call the energy landscape, stems, in
part, from a well-known argument of Cyrus Levinthal~\cite{levinthal}.
The argument starts by noticing that number of possible conformations
in a protein scales exponentially with the number of amino acid
residues.  Thus, if each amino acid has only two possible
conformations, then the number of possible conformations for a protein
with 100 amino acid is $2^{100} \approx 10^{30}$.  If, as a
conservative estimate, at least one picosecond is required to explore
each conformation, then the time required to explore all conformations
of the 100 amino acid protein is approximately $10^{18}$ seconds, or
more than $10^{10}$ {\em years}. From this estimate Levinthal argued
that the protein did not have enough time to find its global free
energy minimum, so the final, folded conformation of a protein must be
determined by kinetic pathways.  This argument is easily criticized.
For example, one could equally well apply it to the formation of
crystals, and conclude that crystallization can never occur!  More
seriously, the argument can be used to question how the protein could
reliably find {\em any} particular conformation.  In this form the
argument is often called Levinthal's paradox.  The weak point in
Levinthal's argument is the assumption that all conformations are
equally likely in the path from the unfolded to the folded states.  In
fact, conformations with lower free energy are more likely than those
with higher free energy.  Levinthal's argument assumes a free energy
landscape that looks like a flat golf course with a single hole at the
free energy minimum.  The argument breaks down completely for a free
energy landscape that looks like a
funnel~\cite{bw_jpc,zwanzig,leopold,dill_cosb}. A central purpose of
this paper is to further develop this intuitive notion of energy
landscape and to describe quantitatively kinetic behavior on the kinds
of energy landscapes that are encountered in protein folding.
Interestingly, Levinthal's paradox will reoccur, albeit in a
completely different form.

The most detailed description of the energy landscape of a folding
protein molecule would be obtained by specifying the free energy
averaged over the solvent coordinates as a function of the coordinates
of every atom in the protein.  At this fine level of description, the
free energy surface of a protein is riddled with many local
minima~\cite{mccammon_harvey,hans_f}. Most of these minima correspond
to small excitation energies connected with individual local
conformational changes such as rotations of individual side chains.
The energies involved in these small conformational changes are
typically on the order of $ k_B T $, that is, the size of the thermal
energies of the atoms in the protein.  Interconversion between these
shallow local minima will be rapid on the time scale of protein
motions.  Sometimes many sidechains can shift, giving quite different
minima with a large energy barrier between them.  Changes of backbone
conformation can lead to globally different protein folds involving
many different inter-residue contacts.  The energies involved in these
larger conformational differences can easily become many times $ k_B T
$, and interconversion between these deeper, globally different local
minima can be quite slow~\cite{mccammon_harvey,hans_f}.

\begin{figure}[tbp]
\insfig{.8}{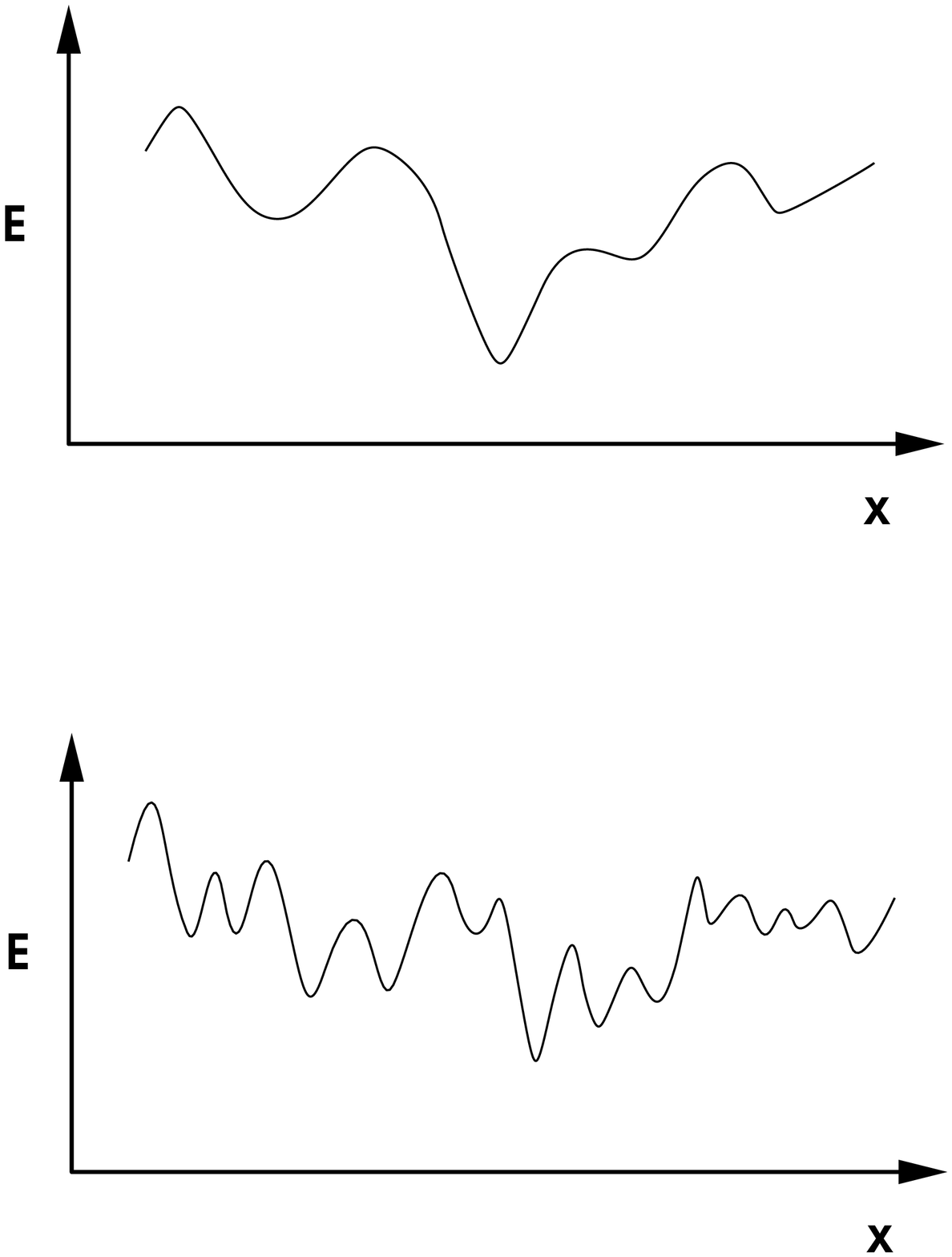}
\caption{The energy of a system (vertical axis) is sketched against a
  single coordinate (horizontal axis) for representative smooth and
  rough energy landscapes.  The top sketch shows a smooth landscape
  with only a few energy minima each having a broad funnel leading to
  it.  The bottom sketch shows a rough energy landscape with many
  energy minima each with a narrow funnel leading to it.  }
\label{fig:englandscape}
\end{figure}

The interesting features of protein folding dynamics concern the free
energy surface viewed on this more coarse-grained structural scale.
Very different behavior occurs, depending on whether this
coarse-grained energy landscape is ``smooth'' or ``rough.''  In
Figure~\ref{fig:englandscape} we show representative smooth and rough
energy landscapes.  A smooth energy landscape has only a small number
of deep valleys and/or high hills.  For smoother energy landscapes
there are typically many high energy structures and only a few low
energy structures.  The more closely the system resembles a few low
energy structures, the lower the energy.  Thus, each of the low energy
structures is at the bottom of a broad energy valley.  A protein
molecule that was in one of the valleys would find itself dynamically
funneled to the lowest energy state.  Therefore, we will refer to the
valley associated with a low energy structure a ``funnel''.  In this
language, a system with a smooth energy landscape has a few deep
minima, each having a large, broad funnel.  Systems with smooth
landscapes exhibit cooperative phase transitions, illustrated by such
phenomena as crystallization of simple materials and in biological
macromolecules by phenomena such as the helix-coil
transition~\cite{poland_scheraga}.  The thermodynamic phases of
systems as smooth energy landscapes are determined by the temperature.
At high temperatures, the large number of high energy structures
predominate, but as the temperature of the system is lowered, the
system will occupy the lower energy states.  Dynamically, below a
transition temperature, such systems will fall into a funnel of low
energy states and may remain trapped there.  In typical cooperative
transitions such as crystallization, once a large enough nucleus of
low energy structure is formed, the rest of the low energy structure
forms rapidly~\cite{ll_statphys1,lp_physkin,ma}.

Thermodynamically, protein tertiary structure formation for smaller
proteins has been shown to exhibit this type of cooperative behavior.
For small, single domain proteins, at most two states are observed on
the longest time scales under physiological solvent conditions: One a
high entropy high energy disordered phase corresponding to the
unfolded protein, and a lower entropy low energy phase describing the
folded protein~\cite{privalov}.  The fact that the phase space is
divided into two main parts is confirmed by the coincidence of
transitions measured by different probes such as optical rotation or
fluorescence~\cite{ginsburg,anfinsen_bj,nojima}.  In addition on the
longest time scales, one sees only a single exponential in the
kinetics of folding.  Simulations of protein folding have shown
evidence of nucleation-like behavior~\cite{dave_1}.  Thus these
aspects of tertiary structure formation are characteristic of a system
with smooth energy landscape.

Smooth energy landscapes are so commonly used in the description of
problems that systems with rough energy landscapes are considered
exotic and have only recently been studied by chemists and physicists.
A rough energy landscape would be one that when coarse grained has
many deep valleys and very high barriers between them. In such a rough
energy landscape there are very large numbers of low energy structures
that are entirely different globally.  Each of these diverse low
energy structures has a small funnel leading to it.

The thermodynamic and kinetic behavior of systems having rough energy
landscapes is quite distinct from those with smooth landscapes.  Rough
energy landscapes occur in problems in which there are many competing
interactions in the energy function.  This competition is called
``frustration.'' The paradigm for a frustrated system is the spin
glass, a magnetic system in which spins are randomly arrayed in a
dilute alloy~\cite{binder_young,mezard,fischer_hertz}.  The
interactions between spins are equally often, at random ferromagnetic
(the spins want to point in the same direction) and antiferromagnetic
(the spins want to point in opposite directions).  These two
conflicting local tendencies (one to parallel spins, the other to
alternating spins) can not be satisfied completely in any arrangement
of spin orientations.  Thus, the system is said to be
``frustrated''~\cite{anderson}. Many optimization problems that arise
in economic contexts have rough energy landscapes because of
frustrated interactions. An economic example of a rough landscape is
provided by the traveling salesman problem.  In this problem one
attempts to minimize the total length of a journey which visits each
of a set of randomly arrayed cities precisely once during the trip.
Here searching for the minimum length trips leads to an optimization
problem in which there are many alternate routes that have very nearly
the same value of the required length (equivalent to multiple minima).
The frustration here arises from the constraint of a single visit to a
city because of an occupancy tax; no central location can be used as a
base.  Finding the optimal solutions to this problem is a difficult
task.  Computer scientists have developed a set of ideas that
describes many problems that are hard to solve~\cite{garey_johnson}.
Although the precise technical framework of these ideas is elaborate,
the basic idea is simple; there exists a set of difficult problems
that can not be solved by any known polynomial time algorithm, and it
is generally believed that no such algorithm exists.  These problems
are called NP-complete.  Here by polynomial time algorithm we mean
that the amount of computation time required to solve the problem
grows no faster than some fixed power of the problem size, {\em e.g.},
the number of cities in the traveling salesman problem.  Furthermore,
the general model of computation used in NP-completeness proofs is
thought to be able to simulate any natural system, so the limitations
that NP-completeness impose on computation probably hold for all
natural systems, {\em e.g.}, folding proteins, the human brain, {\em
  etc.}.  Thus, solutions to NP-complete problems require an
exponential, rather than polynomial, amount of time.  In practical
terms, NP-completeness means that the amount of time required to solve
even modest size problems can become astronomically large.  The
traveling salesman problem an example of a NP complete problem; that
is, its solution for the general case requires exponentially more
computational time as the size of the problem grows.  Finding the
lowest free energy state of a macromolecule with a general sequence
also been shown to be NP-complete~\cite{unger_moult}.  NP-completeness
is a worst case analysis; if a problem is proven to be NP-complete
then finding the solution to at least one case requires an exponential
computation time.  In economic situations these computational
difficulties are avoided by choosing to be satisfied with an
acceptable solution {\em or} by selecting the conditions of the
problem so that easy answers can be found. An example of the latter is
the introduction of the ``hub'' system to airline traffic.  A central
city, perhaps not usually visited, is introduced as a place that can
be multiply visited at little cost.  Similarly for the physicist's
spin glass, there are some specifically chosen arrangements of
ferromagnetic and antiferromagnetic interactions so that each
interaction can be satisfied in a single configuration.  The
arrangements of interactions which do this are relatively improbable.
Therefore, in the context of proteins, NP-completeness means that
there are amino acid sequences that can {\em not} be folded to their
global free energy minimum in a reasonable time either by computer or
by the special algorithm used by Nature.  Thus, in analogy with the
economic situation, either naturally occurring proteins fold to a
structure that is not a global minimum {\em or} they have been
selected to be members of the subset of amino acid sequences that {\em
  can } fold to their global free energy minimum in a reasonable time.
The NP-completeness proof does alone not distinguish between these two
possibilities.  If the latter possibility is correct then one approach
to predicting structure is simulated annealing~\cite{sim_annealing}.
Starting at high temperatures, the system is slowly brought to low
temperature while following its dynamics.  These stochastic search
algorithms parallel the Levinthal paradox for protein folding
kinetics.  Such an approach can only work if the computers energy
landscape is sufficiently close to the one that Nature used.

In any case, even if proteins fold to a structure that is not a global
minimum, {\em i.e.}, folding is kinetically controlled, they must {\em
  reliably} fold to a single structure.  Recent experiments on random
and designed amino acid sequences have shown that reliable folding is
not a universal property of polypeptide chains, and that multiple
folded structures are the rule rather that the
exception~\cite{labean,davidson}. Thus, both theory and experimental
evidence indicate that such reliable folding is characterizes only a
small fraction of amino acid sequences.  Proteins are a subset of this
fraction of reliable folders.  Later in this paper we discuss a
property we call minimal frustration.  Evidence from theory and from
simulation indicates that amino acid sequences with minimal
frustration are likely to fold reliably.

\begin{figure}[tbp]
\insfig{.5}{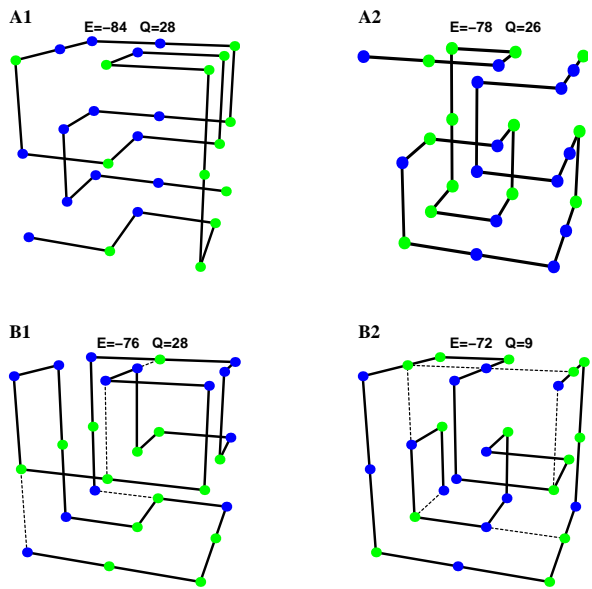}
\caption{The ground state of two different sequences for a $27$-mer,
  with two different types of monomers (two letter code) on a cubic
  lattice.  If two monomers are adjacent in space, but not along the
  chain, then there is an attractive interaction between them.  This
  interaction is strong if the monomers are of the same type and weak
  if they are of different types.  For all figures we use the
  following notation: solid lines represent covalent bonds, dashed
  lines represent spatial contacts with weak interactions and no lines
  are drawn for spatial contacts with strong interactions.  The model
  for this $27$-mer is presented fully in section 6.  The strong
  interactions are equal to $-3$ and the weak ones to $-1$ in an
  arbitrary energy units. The most compact configurations will be
  cube-like and they have $28$ spatial (non-bonded) contacts.
  Sequence (A) has only strong contacts in its ground state.  For this
  reason we call it a non-frustrated ground state.  Figure A1 shows
  the ground state structure for this sequence. We call it
  non-frustrated because all contacts are optimal.  We show in Section
  6 that this sequence is a good folding sequence.  This is not the
  case for sequence (B).  Its ground state configuration has $4$ weak
  interactions, as shown in Figure B1.  For this reason we say that
  this sequence is frustrated, {\em i.e.}, it is unable to optimize
  all the interactions and it has to compromise with some weak ones.
  We show in section 6 that sequence (B) is not a good folder.
  However, there is a more interesting way of observing frustration.
  Let us call $Q$ a measure of similarity between the ground state
  configuration and any other configuration (compact or non-compact)
  for a given sequence.  The quantity $Q$ measures the number of
  contacts between pair of residues that are the same for a given
  configuration and its ground state one.  Therefore, $Q$ is a number
  between $0$ and $28$.  Most of the configurations with energy just
  above the ground state in sequence (A) have $Q$ between $18$ and
  $26$, {\em i.e.}, very similar to the ground state configuration.
  An example of such a configuration is shown in Figure A2 where the
  energy is $-78$ and $Q$ is $26$.  The situation is completely
  different for sequence (B).  There are configurations with energy
  just above the ground state configuration that have a $Q$ between
  $4$ and $12$, {\em i.e.}, they are very different from the ground
  state.  An example of one of these configurations is shown in Figure
  B2 where the energy is $-72$ and $Q$ is $9$.  In this case, there
  are lots of low energy states that are completely different but
  energetically very similar.  When the system gets trapped in one of
  these low energy states, it takes a long time to completely
  reconfigure before it can try to fold again.}
\label{fig:frust}
\end{figure}

What are the possible sources of frustration in the general case of a
heteropolymer?  Consider the hydrophobic effect, which for
illustration we think of as a contact interaction favoring hydrophobic
pairs or hydrophilic pairs.  Because of the constraint of chain
connectivity for most random sequences bringing together a hydrophobic
pair distant in sequence will require bringing together other pairs in
the sequence which will often be dissimilar and therefore
unfavorable.\footnote{It is useful for the reader to study
  Figure~\ref{fig:frust}, in which we illustrate the varying degrees
  of frustration for two sequences of a lattice model of a
  heteropolymer.} This situation could be avoided in natural proteins
by choosing simple patterns of hydrophobic-hydrophilic alternation
like those seen in $\beta$-barrel proteins~\cite{branden_p62}.
Similarly, for most sequences the hydrophobicity pattern favoring a
particular secondary structure ($\alpha$-helix or $\beta$-sheet) might
or might not be consistent with the tendency of each amino acid to be
in that secondary structure.  Indeed, in general is usually some
conflict of this sort, since the ends of $\alpha$-helices have
unsatisfied hydrogen bonds, but the helices must be broken so that a
compact structure can form, satisfying the hydrophobic forces.
Sequences may need special start or stop residues to form terminal
hydrogen bonds gracefully, using sidechains~\cite{presta,richardson}.

Polymers can also exhibit another kind of frustration.  A molecule
often needs to overcome an energy barrier to change from one structure
to another.  This notion has been used explicitly in the simulation
studies of Camacho and Thirumalai and of Chan and Dill where they
constructed paths with minimal energy barriers between similar
configurations in their protein folding models and used this network
of pathways to map out several features of the energy
landscape~\cite{dave_4,chan_dill_tst,chan_dill_homo}. If this energy
barrier is too high to overcome in a reasonable time, for example,
some fraction of the folding time for a protein, then we may say that
the two structures are not ``dynamically connected''.  Two different
structures may resemble each other, and even have similar free
energies, but they may be unable reconfigure from one to the other one
in any reasonable time scale.  Such structures would not be
dynamically connected.  In particular, for polymers, geometrical
constraints arise because the polymer chain can not pass through
itself.  This effect is called excluded volume, and may give rise to
an enormous energy barrier. In this case one can easily have two
structures that resemble each other but are not dynamically connected.
Leopold, Montal and Onuchic have explicitly shown that this situation
occurs in some simple models of protein folding~\cite{leopold}.  We
will refer to this kinetic phenomenon as geometrical frustration.

Systems with rough energy landscapes also exhibit effective phase
transitions~\cite{binder_young,mezard,fischer_hertz}.  When the
temperature of such a system is lowered, it tends to occupy the lower
energy states and at a transition temperature will become trapped in
one of them.  Generally, these transitions are accompanied by a
considerable slowing of the motion as the system tries to exit over
the high energy barriers.  In the case of liquids being super--cooled
below their freezing point, this phenomenon is known as the glass
transition~\cite{brawer,jackle}.  Below the glass transition
temperature, the liquid is trapped in a single deep minimum and thus
it looks like a solid. The thermodynamics of this solid depends on its
detailed thermal history.  Typically, systems with rough energy
landscapes exhibit glass transitions analogous to those that occur
when liquids are supercooled below their freezing point. As the system
approaches the glass transition, the slow transitions between minima
leads to strongly non--exponential time dependence for many
properties.

Typically a heteropolymer with a random sequence interacting with
itself has a rough energy landscape. One source of the roughness is
the frustration arising from conflicting interactions but geometrical
constraints may be important too.  In either case, energetic or
geometrical frustration, there will be a large barrier to reconfigure
between these configurations.  This is a natural starting assumption
for thinking about heteropolymer dynamics since one expects this
behavior generically for heterogeneous systems.  The implications of
the roughness of heteropolymer energy landscapes for protein folding
were first discussed by Bryngelson and Wolynes who postulated that the
energies of the states of a random heteropolymer could be
approximately modeled by a set of random, independent
energies~\cite{bw_pnas}. This model is known as the random energy
model in the theory of spin
glasses~\cite{derrida_prl,derrida_prb,gross_mezard}.  The random
energy model approximation used by Bryngelson and Wolynes was later
shown to be equivalent to a more conventional replica mean field
approximation by Garel and Orland ~\cite{garel_orland} and by
Shakhnovich and Gutin~\cite{sg_bpchem}.  A direct demonstration of the
roughness of the energy landscape for heteropolymers has been carried
out for small lattice model proteins.  Here the exact enumeration of
configurations can be carried out and with simple interactions, it can
be directly established that there are configurations very close in
energy to the ground state that have topologically distinct folds for
most random
sequences~\cite{lau_dill_macro,lau_dill_pnas,chan_dill_sss,yue_dill,sg_nature,sg_jcp}.
Work on realistic lattice models for small proteins confined to their
proper shape (where complete enumeration can be carried out) suggests
the possibility of deep low energy structures that are globally
different in form~\cite{covell_jernigan,hinds_levitt}.  Even for a
well designed sequence ({\em i.e.} one designed to have a smooth
energy landscape) some roughness may remain.  Early direct evidence
for roughness in the energy landscape of protein folding simulations
of designed sequences is provided by the work of Honeycutt and
Thirumalai, which looked for and found deep multiple energy minima in
their simulations of $\beta$-barrel folding~\cite{dave_2,dave_3}.
Finally, the historical difficulty of predicting protein structure
from sequence arises from the ``multiple minimum problem,'' that is,
the existence of many minima in the empirical potential energy
functions used to predict these structures.  The large number of
minima indicates that the energy landscape of these potential
functions is rough.  The importance of the multiple minimum problem,
and therefore the roughness of the energy landscape, as an impediment
to structure prediction has been emphasized by Scheraga and
collaborators~\cite{scheraga}.

Some experimental features of protein folding suggest a considerable
roughness to the energy landscape.  Although protein folding appears
to be exponential in time, short time scale measurements show the
existence of intermediates. Also, multi-exponential decay of
relaxation properties are seen in these early events~\cite{cm_jones}.
Many of the time scales involved in protein unfolding have very large
apparent activation energies, suggesting high energy barriers.  There
is the occasional report of history dependence to protein folding;
although, this is absent from studies on smaller proteins {\em in
  vitro}~\cite{privalov}.

\section{Quantitative Aspects of the Statistics and
  Thermodynamics of A Folding Protein.}

In the previous section we found that a folding protein exhibits
behaviors that are characteristic of both smooth and rough energy
landscapes.  Thus, from the phenomenological viewpoint it is evident
that protein folding occurs on an energy landscape that is
intermediate between the most smooth and the most rough.  A simple
model proposed by Bryngelson and Wolynes interpolates between the two
limits and illustrates the basic ideas of the energy\footnote{In this
  section we use the word ``energy'' to describe the free energy of a
  given complete configuration of the protein.  such a configuration
  has many solvent configurations consistent with it.  Thus our energy
  landscape has a temperature dependence due to hydrophobic forces.
  We do not consider this effect when we discuss the pure effects of
  temperature in this section.} landscape analysis of protein
folding~\cite{bw_pnas}. When stripped down to its bare essentials,
this picture of the folding landscape is based on two postulates: The
first captures the rough aspects of the energy landscape.  It is
postulated that (for natural proteins) the energy of a contact between
two residues which does not occur in the final native structure of a
protein or the energy of a residue in a secondary structure which does
not turn out to be ultimately correct can be taken as random
variables; that is, in its non-native interactions, a protein
resembles a random heteropolymer.  In its extreme form this suggests
that we can take the energies of globally distinct states to be random
variables which are uncorrelated, provided no native contacts are made
and no native secondary structure is formed.  A second postulate
captures the smooth aspects of the folding landscape.  When a part of
the protein molecule is in its correct secondary structure, the energy
contributions are expected to be stabilizing.  In addition, when a
correct contact is made, although occasionally the energy may go up,
on the average over all possible contacts, the energy will go down.
Thus if the similarity to the native structure is used as a distance
measure, the surface may have bumps and wiggles but the energy
generally rises as we move away from the native structure.  Thus there
is an overall energetic funnel (of the sort discussed in the previous
section) to the native structure.

Bryngelson and Wolynes used the term the principle of minimal
frustration in describing the smoothness postulate, insofar as it is
what distinguishes natural proteins from random heteropolymers.  The
smoothness of folding landscapes arises from the selection of protein
sequences by evolution.  If the necessity to maximize the ability of
folding quickly is the dominant selection pressure, the smooth part of
the energy landscape will be paramount.  On the other hand, there are
other selection pressures as well.  Thus evolution may not be able to
remove some frustrated interaction from natural proteins.  Indeed,
neutral evolution would suggest that randomness and frustration would
continue to exist to an extent that allows only adequate stability and
kinetic foldability.  The minimal frustration of natural proteins is
evident in several ways.  Examination of X-ray structures shows that
sidechains are in fact chosen by evolution to make coherent
contributions to supersecondary structures.  Clear examples are
leucine zippers ~\cite{zipper} and the $\beta$-barrel amphiphilicity
mentioned earlier.  Symmetric sequences like these often lead to low
frustration in symmetric structures.  Consistency between secondary
structures and global tertiary structures is also important.  This is
the ``principle of structural consistency'' enunciated by
G\={o}~\cite{go_annrev_biophys}.

Purely kinetic effects also limit the folding of proteins.  For
example, if the minimum energy structure is not dynamically connected
(in the sense described in the previous section) to any other low
energy structures, then it would be kinetically inaccessible in spite
of its low energy.  The importance of kinetic effects for protein
folding was investigated in the previously mentioned study of Leopold,
Montal, and Onuchic~\cite{leopold}.  They simulated the folding of two
``sequences'' with their simple model, one of which folded rapidly to
its global energy minimum, the other of which failed to find its
global energy minimum in several long runs.  Analysis of the dynamical
connectivities produced by the two ``sequences'' showed that the
minimum energy structure of the rapidly folding sequence had a rich
network of dynamical connections to most of the other low energy
structures.  In contrast, the minimum energy structure of the other
sequence was sparsely connected to other low energy structures.

\begin{figure}
\insfig{.6}{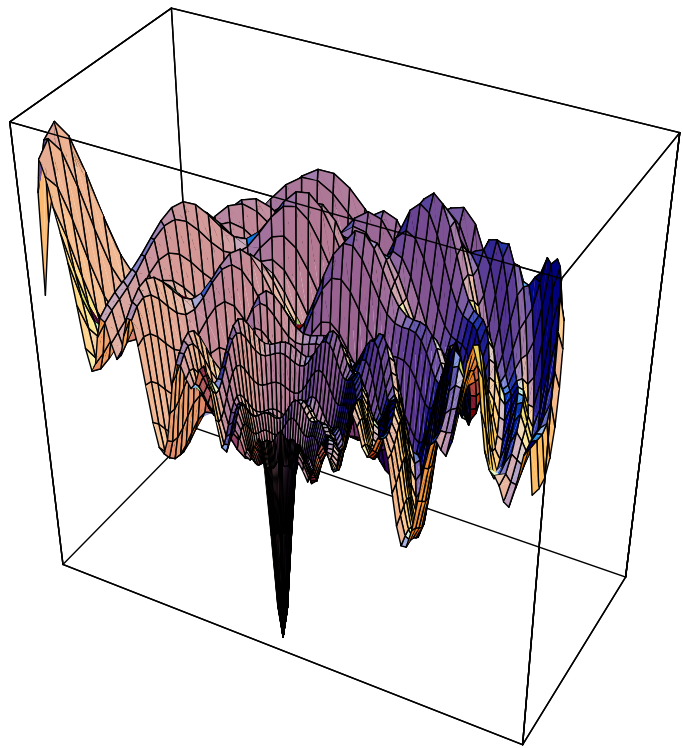}
\caption{\hskip-3pt(a)
  Sketch of an energy landscape encompassing the qualitative
  considerations about the folding.  The energy is on the vertical
  axis and the other axes represent conformation.  This landscape has
  both smooth and rough aspects.  Overall, there is a broad, smooth
  funnel leading to the native state, but there is also some roughness
  superimposed on this funnel. Of course, no low dimensional figure
  can do justice to the high dimensionality of the configuration space
  of a protein.}
\label{fig:landscape}
\end{figure}

A figure encompassing the qualitative considerations about the folding
landscape is pictured in Figure~\ref{fig:landscape}a.  Of course, no
low dimensional figure can do justice to the high dimensionality of
the configuration space of a protein, but one sees that the dominant
smooth features of the landscape depend on how close a protein
configuration is to the native one and this coordinate is specifically
shown as the radial coordinate in the figure.  There are a variety of
ways of measuring the similarity of a protein structure to the native
structure.  One can take the fraction of the amino acids residues
which are in the correct local configuration.  This is a choice used
in the original papers of Bryngelson and
Wolynes~\cite{bw_pnas,bw_jpc,bw_biopolymers}.  Another possibility for
measuring tertiary structure is the fraction of pairs of amino acids
which are correctly situated to some accuracy.  This measure is
related to the distance plots used by
crystallographers~\cite{schulz_schirmer,creighton_proteins}. The
similarity measure may also be thought of as a measure of the distance
between the two structures, so that similar structures are considered
to be close to one another.  We denote the similarity of a protein
structure to the native structure by $ n $.  We will take $n=1$ to
denote complete similarity to the native state and $n=0$ to denote no
similarity to the native state.  The radial coordinate in
Figure~\ref{fig:landscape}a should be thought of as this similarity
measure $n$.  The average energy of state with a certain similarity to
the native structure has a value that gets lower as the native
structure is approached - thus the overall slope of the energetic
funnel.  On the other hand the rugged part of the energy landscape
means that no individual state has precisely this energy and we can
characterize the fluctuations in energy with a given similarity to the
native structure by the variance, $\Delta E^2 (n)$.  The ruggedness of
the energy landscape as measured by this variance clearly depends on
the compactness of the protein molecule since it arises from improper
3-dimensional contacts.  In general, the variance may also conceivably
decrease as the native structure is approached, but this is not
essential for our picture.

The energy of a given state arises from the contributions of many
terms, so it is natural to assume that the probability distribution of
energies for any similarity to the native structure is given by a
Gaussian distribution,
\begin{equation}
  P(E) = { 1 \over \sqrt{2 \pi \Delta E^2}(n)} \exp -{ \left(E - \bar
    E (n) \right)^2 \over 2 \Delta E^2(n)}.
\end{equation}

The other important feature of the statistical landscape description
is the number of conformational states of protein as we move away from
the native structure.  The total number of conformational states grows
exponentially with the length of the protein.  If there are $\gamma$
configurations per residue, this total number of configurations is
$\Omega = \gamma^N$. $\gamma$ depends on the level of description of
the model.  It is of order 3, 4 or 5 for the backbone coordinates, but
might rise to roughly 10 if the sidechain configurations are also
included in the analysis.  As noted above, the ruggedness of the
energy landscape is most important when the protein is compact.  The
number of compact configurations is considerably smaller than the
total number and can be estimated from Flory's theory of excluded
volume in polymers~\cite{flory_jcp,flory_book}, $\Omega(R)$ decreases
quite considerably as the radius of gyration of the protein falls.
For maximally compact configurations of the backbone, $\Omega(R)^* =
\gamma^{*N}$ where $\gamma^*$ is of the order 1.5.

The completely folded protein has a much smaller degree of
conformational freedom.  Essentially a single backbone structure
exists.  Thus the number of configurations of the protein decreases as
we move toward the native structure.  Therefore, if $\Omega(n)$
denotes the number of structures with a similarity measure with the
native structure of $n$, then $\Omega(n)$, and
\begin{equation}
  S_0(n) = k_B log \Omega(n)
\end{equation}
decreases as $n$ gets larger.  The exact similarity measure determines
the behavior of $S_0(n)$.  For our purposes here we need only take a
simple form of $S_0(n)$ that decreases as the native state is
approached.  Roughly speaking, we can approximate $\Omega$ as a
function of $n$ by $\Omega = \gamma^{*N(1 - n)}$.

\addtocounter{figure}{-1}

\begin{figure}
\insfig{.8}{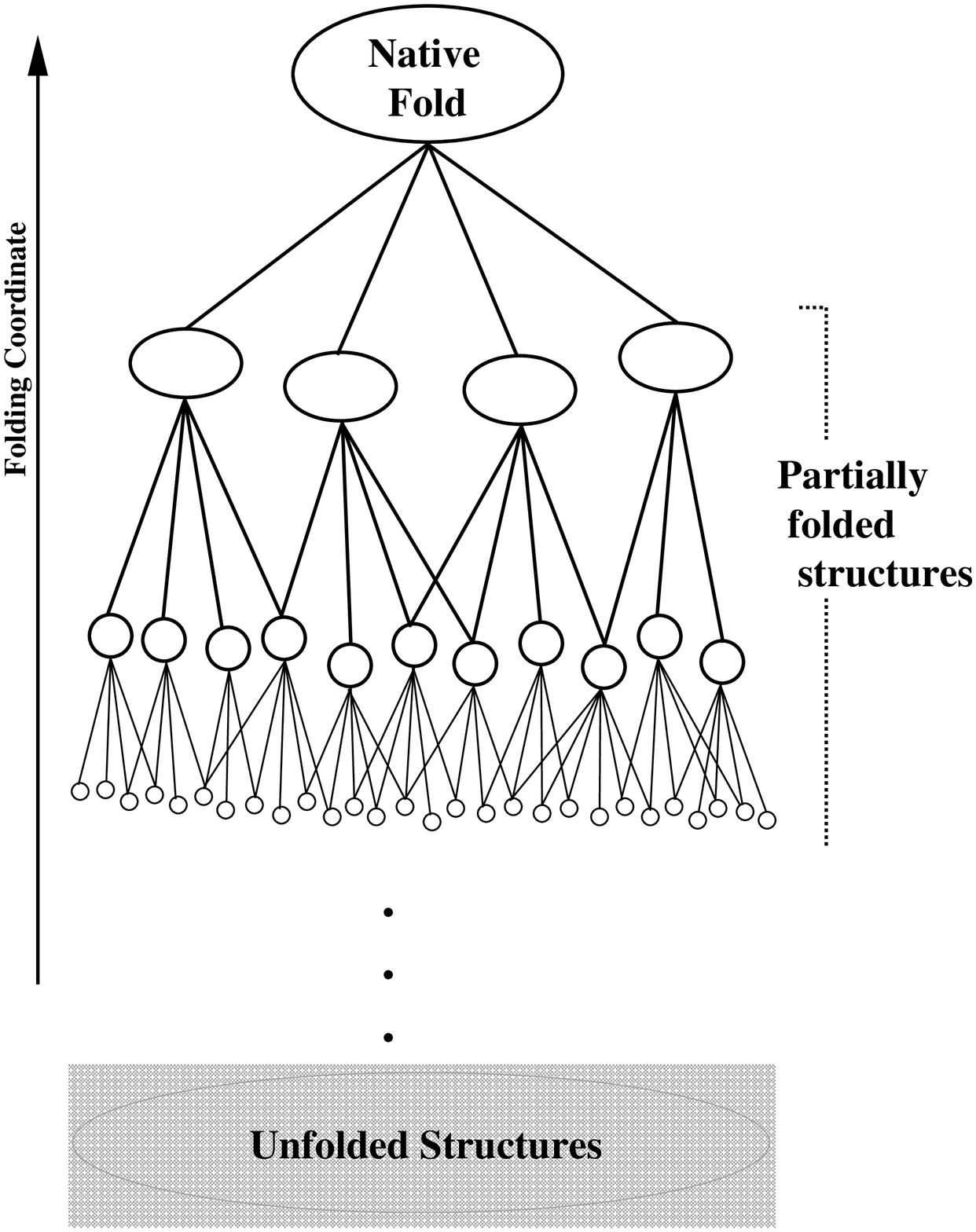}
  \caption{\hskip-3pt(b)
    A schematic drawing of protein conformations in relation to their
    similarity to the native state.  The vertical direction is a
    folding reaction coordinate.  The conformations that are higher in
    the figure are more similar to the native state.  As one moves
    away from the native structure there is a huge increase in the
    number of possible conformations.}
\label{fig:schematic}
\end{figure}

As one moves away from the native structure there is a huge increase
in the number of accessible states, which we can think of as living on
the branches of a highly arborized tree as is shown schematically in
Figure~\ref{fig:schematic}b.  Not all of the states on a statistical
landscape are thermodynamically or kinetically important, since the
high energy states cannot be thermally occupied.  The number of states
with a specified energy $E$, which have a specified similarity, $n$,
to the native structure, is given by
\begin{equation}
  \Omega(E,n) = {\gamma^{*}}^{N(1 - n )} P(E) \ .
\end{equation}

At thermal equilibrium, only a small band of energy is occupied with a
certain similarity to the native structure.  For a large protein, this
band will be relatively well defined in energy.  The most probable
value of the energy in this band can be obtained by maximizing the
thermodynamic weight of states of a given energy.  This is a product
of the Boltzmann factor and the number of states of that energy
\begin{equation}
  p(E,n) = { 1 \over Z} \Omega(E,n) e^{-E / k_B T} \ .
\end{equation}
(Note: Do not confuse $p(E)$ above with the $P(E)$ defined in equation
1.)  Here $Z$ is the partition function, which ensures normalization
of the probability function.  Thus the most probable energy with a
certain similarity to the native structure is given by
\begin{equation}
  E_{m.p.}(n) = \bar E(n) - {\Delta E(n)^2 \over k_B T} \ ,
\label{energy_mp}
\end{equation}
and the number of thermally occupied states is
\begin{equation}
  \Omega(E_{m.p.}(n),n) = \exp \left[ { S_0(n) \over k_B} - {\Delta
    E(n)^2 \over 2 ( k_B T)^2} \right]
\label{omega_mp}
\end{equation}
The entropy of the thermally occupied structures that have a certain
similarity to the native structure is,
\begin{equation}
  S (E_{m.p.}(n), n) = k_B \log (\Omega(E_{m.p.}(n),n)).
\label{entropy_mp}
\end{equation}

We see from these expressions that there are two opposing
thermodynamic forces involved in the folding process.  The growth in
the number of thermally occupied states as we move away from the
native structure favors a large number of highly disordered
configurations.  On the other hand, the decreasing average energy as
one approaches the native structure favors folded configurations.
These two features are combined by thinking of the free energy as a
function of the configurational similarity $n$ at a fixed temperature
$T$,
\begin{eqnarray}
  F(n) &=& E_{m.p.} (n) - T S (E_{m.p.}(n), n) \nonumber \\ &=&
  \overline{E}(n) - \frac{\Delta E(n)^2}{2k_B T} - TS_0(n) \ .
\label{free_energy}
\end{eqnarray}

\begin{figure}
\insfig{.8}{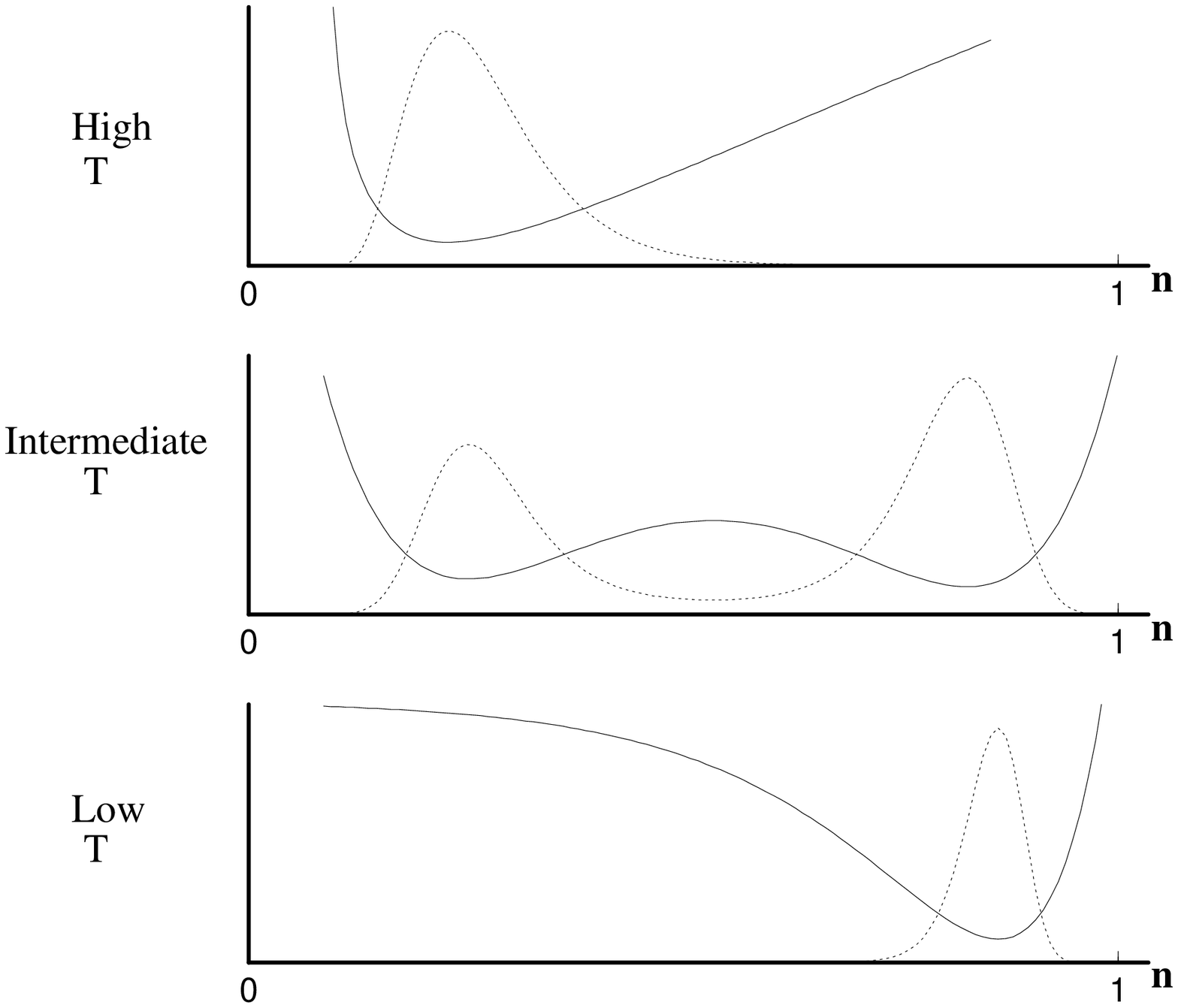}
\caption{Sketches of the free energy
  (solid lines) and probability of occupation (dashed lines) against a
  folding reaction coordinate for three different temperatures.  The
  value $n=1$ corresponds to the native structure.  The top sketch
  shows the situation for high temperatures, where the free energy
  function has a single minimum near $n=0$, i.e., in highly unfolded
  states.  Here the molecule is far more likely to be in an unfolded
  conformation than it is to be a conformation similar to the native
  structure, as it is shown by the dashed lines.  As the temperature
  is lowered, the free energy develops a second minimum, one of them
  similar to native structure.  There is a a free energy barrier
  between these minima.  At these temperatures the probability of
  occupation is bimodal, with one unfolded and one nearly native peak.
  Finally at low temperatures, there is again a single minimum in the
  free energy, but this minimum is near the native structure.  Here
  the molecule is very similar to the native structure.}
\label{fig:feprb}
\end{figure}

\noindent
This free energy function is the logarithm of the thermodynamic weight
of states with a certain similarity to the native structure.  We see
in Figure~\ref{fig:feprb} a representation of this free energy and of the
probability of occupation.  At high temperatures, the band of states
with nearly no native structure is favored, corresponding to an
unfolded state.  At very low temperatures the folded configurations
would be favored and in between, a double minimum effective free
energy pertains.  The folding temperature is determined by the
condition that the two global minima be equal in thermodynamic weight.
The unfolded minimum can correspond to two distinct sets of states
corresponding to different values of a distinct order parameter, the
radius of gyration~\cite{bw_biopolymers}.  If the randomness is large
and non-specific interactions are important, or the chain is highly
hydrophobic in composition, this minimum itself can be collapsed.
This may well correspond to the molten globule state~\cite{ptitsyn}.
On the other hand, if there is little average driving force to
collapse due to non-specific contacts ($\Delta E(n)^2$ small) the
disordered configurations will non-compact and this corresponds to the
traditional denatured random coil state.  We note that many
intermediate degrees of order can exist in the molten globule phase
and these can and should be taken into account in a complete analysis.
However, the simple one-parameter analysis captures the essentials and
should fit data over an appropriately restricted range of
thermodynamic conditions.

\section{Quantitative Aspects of the Kinetics of a Folding
  Protein.}

The theoretical formulation of the kinetics of protein folding differs
from the classic formulation of transition state theory in some
important ways.  Most of our ideas concerning rate theory had their
origin in studies of gas phase reactions of small molecules and simple
unimolecular reactions in liquids~\cite{glasstone,moore_pearson}.
Four important properties of these simple reactions will illustrate
the most important points of contrast with protein folding.  First, in
the simple reactions solvent is either absent or plays a passive role,
{\em e.g.}, as a heat bath or source of friction.  Second, the initial
state, final state and transition state all refer to single, fairly
well defined structures so entropy considerations are not important.
Third, there is a single, fairly well defined reaction coordinate.
Fourth, the effective diffusion coefficient for moving along the
reaction coordinate changes very little as the system moves from the
initial to the final state.  Protein folding is completely different
from these simple
reactions~\cite{bw_jpc,dave_4,chan_dill_tst,nature_paper}.
First, in protein folding, the solvent plays a vital role in
stabilizing the folded state.  As explained above, the important role
of the solvent means that the potential of mean force, which here
plays the role of the energy as a function of configuration, is a
function of temperature and solvent conditions.  Second, the initial
denatured state, final folded state and transition states all refer to
sets of protein structures, so the configurational entropy of the
protein chain is a necessary part of the description of protein
folding.  Third, there are many possible reaction coordinates and
pathways.  Fourth, the dynamics of the protein chain changes
qualitatively during the course of folding; in particular, an open
chain, has far greater thermal motion than a collapsed chain.
Therefore, the effective diffusion coefficient for motion along a
reaction coordinate for folding probably can also change qualitatively
between the initial and final states.  Below we will discuss the
modifications to the transition state theory framework that are needed
to describe protein folding kinetics.

The gradient of the free energy function, $F(n)$, describes the
overall tendency for the system to move and change its value of $n$.
The average flow in configuration space will tend to minimize the free
energy.  For typical forms of the expressions for the energy and the
entropy as a function of similarity to the native state, $n$, $F(n)$
will tend to have one or two minima, so the system will be unistable
or bistable.  If the system is unistable and the conditions are
favorable for folding, then the single minimum of the free energy
function must occur near the native state.  A unistable free energy
function with its minimum near the native state would require a huge
thermal driving force.  We call this situation ``downhill'' protein
folding. Downhill folding is rare in slow timescale {\em in vitro}
protein folding experiments carried out in conditions near the
transition between equilibrium folding and equilibrium unfolding.
Downhill folding may be common in strongly nativizing conditions, in
the initial stages studied in fast timescale folding
experiments~\cite{cm_jones} and {\em in vivo}.  In downhill folding
the protein folds by making a straight run down the average free
energy gradient.

An analogy with transition state theory~\cite{glasstone,moore_pearson}
yields a simple estimate for the folding rate, or equivalently, the
folding time~\cite{bw_jpc}.  In transition state theory the reaction
rate is given by the rate of going through the bottleneck for the
reaction.  Traditionally, this bottleneck is the highest free energy
state in the reaction coordinate pathway from the reactant state to
the product state.  This bottleneck is called the transition state.
In transition state theory the rate of going through the transition
state depends on the free energy barrier, {\em i.e.}, the difference
between the transition state free energy and the reactant state free
energy.  In downhill folding there is no free energy barrier.
However, there is a bottleneck for folding in downhill folding,
because the effective diffusion coefficient for motion along a
reaction coordinate changes qualitatively during the course of
folding; the region with the smallest diffusion coefficient is the
kinetic folding bottleneck.  Let $\overline{t}(n)$ denote the typical
lifetime of an individual microstate with a similarity $n$ to the
native structure.  This lifetime is a measure of the rate of motion
along the reaction coordinates for folding; the larger $\overline{t}$
the smaller the effective diffusion coefficient and the slower the
folding rate.  The kinetic bottleneck for folding occurs at the value
of $n$ that maximizes $\overline{t}(n)$, which we denote by
$n_{kin}^{\ddagger}$.  The subscript $kin$ stands for kinetic and the
reason for using this subscript will become apparent below.
Therefore, a simple estimate of the folding time, $\tau$, in analogy
with transition state theory, is given by
\begin{equation}
  \tau = \overline{t}(n_{kin}^{\ddagger}).
\label{downhill_rate}
\end{equation}
Notice that the time in equation (~\ref{downhill_rate}) is a lower
bound on the folding time, hence an upper bound on folding rate.  This
property is is expected because the transition state technique gives
upper bounds on reaction rates~\cite{moore_pearson}.  We shall discuss
the meaning of $n_{kin}^{\ddagger}$ in more detail below.  For now
notice that $n_{kin}^{\ddagger}$ is {\em not} the location of the top
of the free energy barrier, as in conventional transition state
theory.

The roughness of the energy surface determines the lifetime of
individual microstates.  The detailed distribution of these lifetimes
can be determined from a detailed analysis and it is rather broad.
However, a reasonable first approximation to the typical escape time
is easy to obtain.  Most minima along a perimeter of constant $n$ are
surrounded by ordinary states with nearly the average energy, $\bar
E(n)$.  Thus the barrier height for hopping is $\bar E - E_{m.p.} =
(\Delta E/k_B T)^2$.  This gives an escape time with a super-Arrhenius
temperature dependence
\begin{equation}
  \bar t(n) = t_0 e^{(\Delta E(n) / k_B T)^2}
\label{t_bar}
\end{equation}
The prefactor $t_0$ is the timescale for a typical motion of a large
segment of the chain.  It depends on local barriers and on the solvent
viscosity, which is itself temperature dependent.  Isoviscosity
studies of protein folding are therefore quite interesting.  The
non-Arrhenius temperature dependence exhibited here is sometimes
called the Ferry law ~\cite{ferry} and is describes the slow dynamics
of many glassy systems.  As expected, increasing the roughness of the
energy landscape greatly slows down folding.

What happens to the escape process as the temperature is lowered?  The
above estimate assumes many channels for escape exist and an average
one can be taken.  But as the temperature is lowered it becomes
preferable to find an unlikely channel with an improbably low barrier.
A subtle analysis ~\cite{bw_jpc} shows that, for a given value of $n$,
the escape time goes no lower than a ``search'' time
\begin{equation}
  \overline{t} = t_0 e^{S_0(n)/k_B}
\label{search_time}
\end{equation}
This is the average number of steps taken by the protein to find a
state of negligible barrier.  This is the Levinthal time for searching
states at fixed perimeters, i.e., fixed value of $n$.  For a given $n$
this escape time is reached at a temperature
\begin{equation}
  T_g(n) = \left( \frac{\Delta E(n)^2}{2 k_B S_0(n)} \right)^{1/2}
\end{equation}
The analysis of Bryngelson and Wolynes also shows that for $ T>T_g(n)
$ the protein has kinetic access to representative section of the
perimeter (see Fig.~2) so the behavior of a typical protein molecule
can be replaced by the behavior of a statistical ensemble.  In this
case equations (~\ref{downhill_rate}) and (~\ref{t_bar}) for the
folding time are valid.\footnote{The more subtle analysis by
  Bryngelson and Wolynes shows that the full time dependence of
  $\overline{t}$ is slightly more complicated: For $T>2T_g(n)$,
  $\overline{t}(n) = t_0 \exp [(\Delta E(n) /k_B T)^2]$, as in
  equation (\ref{t_bar}) above, but for $2T_g(n) > T > T_g(n)$, this
  equation must be modified to $\overline{t}(n) = t_0 \exp [S_0(n) +
  (1/k_B T_g(n) -1/k_B T)^2 \Delta E(n)^2]$} For $ T<T_g(n) $ the
protein has kinetic access to very few structures.  These structures
are not necessarily representative of the statistical ensemble, so the
proteins behavior is dominated by the details of its specific energy
landscape.  In this case equations (~\ref{downhill_rate}) and
(~\ref{t_bar}) for the folding time must be modified.  Technically,
the kinetic behavior of the protein molecule becomes
non-self-averaging, a term we discuss later in this section.

A system with a fixed $n$ also undergoes a thermodynamic second order
phase transition at $T_g(n)$ in which the protein is effectively
frozen into one or few of a small number of low energy states.  Using
Eq.~6, we see that for $T \le T_g(n) $, the number of thermally
occupied states no longer scales exponentially with the size of the
protein.\footnote{Applying equation 6 literally would imply a
  thermally accessible perimeter with less than one state because the
  entropy analysis neglects finite size corrections.} Conversely, as a
protein folds at a fixed temperature $T$, the similarity to the native
structure, $n$, becomes larger.  However, the entropy, $S(E_{m.p.}(n),
n)$, decreases as $n$ becomes larger, {\em i.e.}, there are fewer
states available to the protein molecule as it approaches its native
structure.  A typical protein runs out of entropy at some value $n_g$
of $n$.  This vanishing of configurational entropy is precisely the
previously noted second order phase transition, this time taking $T$
rather than $n$ to be constant.  The critical values of the
temperature and the fraction native structure are related by
$T=T_g(n_g)$, where $T$ is the temperature at which the folding
experiment is carried out.  In addition, Bryngelson and Wolynes have
shown that the glass transition can only occur for a protein that
already has collapsed~\cite{bw_biopolymers}.  Therefore, for any given
temperature $T$, for values of $n \le n_g(T)$ the kinetic description
presented in this section is valid and the folding kinetics are
self-averaging, and for $n > n_g(T) $ the protein is in the glassy
phase, and its kinetics becomes non-self-averaging.

For bistable systems, there are two minima of free energy with a
maximum of free energy between them.  In folding conditions the
minimum close to the native state has a lower free energy than the
minimum corresponding to the unfolded state.  The free energy barrier
for folding, $ F_{barrier} $, is given by the difference between the
free energy of the unfolded minimum, $F(n_{uf})$ and the free energy
of the the barrier top, $F(n_{th}^{\ddagger})$.  The subscript $th$
stands for thermodynamic and the reason for using it will become clear
momentarily.  Systems to the right of the top of the free energy
barrier, {\em i.e.}, with $n>n_{th}^{\ddagger}$, tend to become
folded; those to the left, {\em i.e.}, with $n<n_{th}^{\ddagger}$,
would become unfolded on the average.  A straightforward
generalization of transition state theory ~\cite{bw_jpc} indicates
that the overall folding time is given by
\begin{equation}
  \tau = \overline{t}(n_{kin}^{\ddagger}) e^{F_{kin}^{\ddagger}/ k_B
    T}
\label{BW_rate}
\end{equation}
where $F_{kin}^{\ddagger} = F(n_{kin}^{\ddagger}) - F(n_{uf})$ and
$n_{kin}^{\ddagger}$ is the value of $n$ that maximizes the above
expression for $\tau$.  One may think of $n_{kin}^{\ddagger}$ as the
similarity to the native state where the bottleneck for folding
occurs. The set of states with $n=n_{kin}^{\ddagger}$ acts like the
transition state for folding when we consider influences of external
agents on rates.

Although equation (~\ref{BW_rate}) for the folding time has the same
form as analogous expressions from traditional transition state
theory, there are three important differences.  First, the prefactor
is $ \overline{t}(n_{kin}^{\ddagger}) $, the typical lifetime of an
individual microstate at a similarity $n_{kin}^{\ddagger}$ to the
native structure.  The corresponding prefactor in absolute rate theory
would be an expression only involving fundamental constants.  The need
for the prefactor based on the lifetime of the microstates stems from
the greater complexity of protein folding as compared with the gas
phase reactions which absolute rate theory was originally designed to
describe.  This lifetime strongly depends on the roughness of the
surface.  Ignoring this fact, we see that the folding is considerably
less than the Levinthal estimate, because some of the configurational
entropy loss is balanced by the gain in energy as the native structure
is approached.  The second difference is that $n_{kin}^{\ddagger}$,
the analogue of the transition state in equation (~\ref{BW_rate}) for
the folding time, is determined by maximizing the {\em entire} folding
time expression in this equation.  In contrast, in traditional
transition state theory, the transition state is a maximum of the free
energy, which would here correspond to $n_{th}^{\ddagger}$.  If the
average lifetimes $\overline{t}(n)$ were constant, {\em i.e.},
independent of $n$, then $n_{kin}^{\ddagger}$, would equal
$n_{th}^{\ddagger}$.  However, in protein folding, we expect the
average lifetimes $\overline{t}(n)$ to vary strongly with $n$, so
$n_{kin}^{\ddagger}$ will {\em not} always equal $n_{th}^{\ddagger}$
and the difference can be large and important.  More concisely, the
position of the {\em kinetic} folding bottleneck,
$n_{kin}^{\ddagger}$, is not necessarily the same as the position of
the {\em thermodynamic} folding bottleneck, $n_{th}^{\ddagger}$.
Third, whereas in traditional transition state theory the transition
state typically is a specific configuration, the transition state in
our folding time expression (~\ref{BW_rate}) corresponds to an entire
band of states in the full configuration space and should not be
thought of as a unique configuration.  Furthermore, since the
potential of mean force of the protein chain is dependent on
temperature and solvent conditions, the location of the transition
state band will change as the temperature and solvent conditions
change.  This situation is in marked contrast to the case of small
molecules in the gas phase in which the transition state can be
thought of as a single structure which is fixed for all reaction
conditions.

Notice that the free energy gradient provided by the minimal
frustration principle leads to multiple paths approaching this
transition state surface as long as the glass transition has not been
reached and that this is crucial to overcoming the entropy loss on
folding.  The expected temperature dependence of the folding time is
obtained by combining equation (~\ref{BW_rate}) for the folding time
and equation (~\ref{t_bar}) for the average lifetime of a microstate.
The result, after taking the logarithms in order to simplify the
resultant expressions, is
\begin{equation}
  \log \left( \frac{\tau}{t_0} \right) = \frac{F_{kin}^{\ddagger}}{k_B
    T} + \frac{\Delta E(n_{kin}^{\ddagger})^2}{(k_B T)^2}
\label{arrhenius1}
\end{equation}
Notice that if $F_{kin}^{\ddagger}$ and $\Delta E(n_{kin}^{\ddagger})$
are assumed to be temperature independent, then equation
(~\ref{arrhenius1}) implies that an Arrhenius plot of folding time
versus inverse temperature would be curved, and in fact parabolic.
Such curved Arrhenius plots are frequently observed in protein folding
experiments~\cite{creighton_pain_book}.  Unfortunately, these plots
can not be used to derive values for $F_{kin}^{\ddagger}$ and $\Delta
E(n_{kin}^{\ddagger})$ directly.  First, our discussion of microstate
lifetimes is rather rough.  A more careful treatment shows that the
exponent in the expression for the lifetimes (~\ref{t_bar}) must be
replaced with a general quadratic in $\Delta E(n)/k_B T$ when the
system gets close to the glass transition.  Second, and more
important, $F_{kin}^{\ddagger}$ {\em does} depend on temperature,
because the free energies of the unfolded state and the folding
bottleneck, the position of the folding bottleneck, the potential of
mean force of the protein molecule all change with temperature.
Similarly, $\Delta E(n_{kin}^{\ddagger})$ also depends on temperature.
The main point here is that a curved Arrhenius plot of the folding
time should be expected as an elementary consequence of energy
landscape properties of protein folding.

The glass transition, discussed above for downhill folding, also
occurs in systems with bistable free energy functions, in exactly the
same way.  As before, when the system $ T > T_g (n) $ the behavior of
a typical protein can be replaced by the behavior of a statistical
ensemble, so equations (~\ref{BW_rate}) and (~\ref{arrhenius1}) for
the folding time are valid.  For $ T < T_g (n) $ the kinetics are
dominated by the details of the energy landscape, so equations
(~\ref{BW_rate}) and (~\ref{arrhenius1}) for the folding time must be
modified.  The kinetic behavior in this glassy regime is
non-self-averaging, a term we now discuss.

An important feature of protein folding below the glass transition is
non-self-averaging behavior.  The idea of non-self-averaging is best
approached by first discussing its opposite, self-averaging behavior.
In simple terms, a self-averaging property is one that depends on the
overall composition of an object, rather than its detailed structure.
An illustration of this idea is provided by alloys, for example,
brass, an alloy of copper and zinc.  No order determines whether a
particular lattice site is occupied by a copper atom or a zinc atom,
so each piece of brass is different on the atomic scale.  However, in
spite of these differences, all pieces of brass with sensibly the same
composition share many properties, for example, hardness, density,
electrical conductivity, {\em etc.}.  These properties are called
self-averaging because, the value of the property, say hardness, of
member of a statistical ensemble, here pieces of brass with the same
composition, is almost always nearly equal to the the average value of
that property over the statistical ensemble.\footnote{More precisely,
  consider a statistical ensemble of objects, and some property of the
  objects in the ensemble.  A property is called self-averaging if the
  fluctuations of the value of that property in the members of the
  statistical ensemble are small compared to the average value of the
  property over the ensemble.  More detailed discussions of
  self-averaging can be found in the references on spin glasses that
  we cited.} Notice that self-averaging is a characteristic of the
ensemble and the property taken together.  Going back to the alloy
example, density is a self-averaging property for all pieces of brass
with a specified composition, but is not a self-averaging property for
all pieces of metal.  As a biochemical example, consider the ensemble
of amino acid sequences with the same length and amino acid
composition as hen lysozyme.  The ability to form a collapsed globule
with approximately the radius of gyration as a lysozyme molecule is
probably a self-averaging property for this ensemble, whereas the
ability to fold to a structure that hydrolyzes glycosidic bonds is
almost certainly a non-self-averaging property.

The presence or absence of self-averaging of a given property has
important practical implications.  If a property is self-averaging
over some ensemble, then studying that property in one member of the
ensemble suffices to learn about the property for all members of the
ensemble; if the property is non-self-averaging, then studying that
property in one member of the ensemble provides no information about
the property for other members of the ensemble.  The question of
whether or not a given property is self-averaging is also intimately
related to the question of whether or not that property is strongly
affected by mutations.  A mutation will create a new sequence, {\em
  i.e.}, a new member of the ensemble.  A self-averaging property will
behave in the same way in the mutant as in the rest of the members of
the ensemble, but a non-self-averaging property will behave
differently in each member of the ensemble, including the mutant.

In protein folding there are several different ensembles over which
one can average, a few of which we now list, going from the largest,
most general ensemble to the smallest, most specific ensemble.  First,
there is the most general ensemble relevant to protein folding, that
of all possible polymers of amino acids.  Experiments on random
polypeptide sequences explore this ensemble~\cite{labean}.  Next is
the set of ensembles of amino acid sequences with fixed amino acid
composition.  Experiments that investigate random sequences with only
a few types of amino acids have studied instances of these
ensembles~\cite{davidson,rao}. Interestingly, there is some evidence
from computer simulations of protein folding that the collapse time
for a sequence depends only on its composition; this evidence
indicates that collapse time may be a self-averaging property over
these ensembles~\cite{socci}. Finally, there are the ensembles of
sequences that fold to a specific structure, {\em e.g.}, the different
lysozyme sequences mentioned in the introduction.  These ensembles are
studied in research programs that investigate the properties of
different mutants of a particular protein.

\begin{figure}
\insfig{.7}{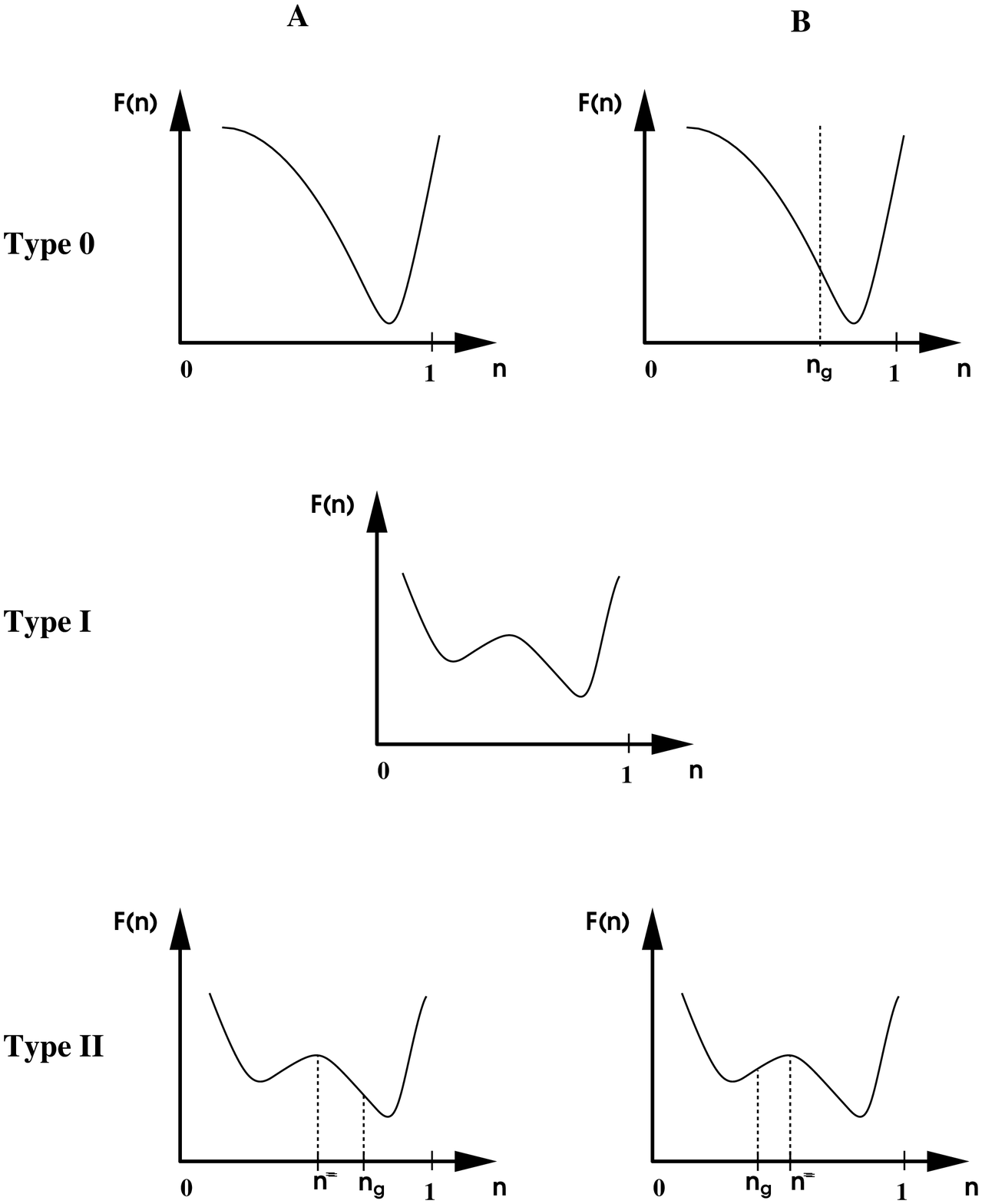}
\caption{Schematic illustrations of the
  folding scenarios discussed in the text.  Each sketch shows a
  qualitative plot of the free energy against the folding coordinate.
  In the type 0 scenarios shown at the top, the free energy function
  has only one minimum near the folded state, i.e., $n=1$.  In a type
  0A transition, shown at the left, there is no glass transition.  In
  a type 0B transition, shown at the right, at some value of the
  folding coordinate, $n_g$, the protein undergoes a glass transition
  and it exhibits the glassy dynamics described in the text for the
  remaining of the folding process, $n>n_g$.  The type I scenario is
  shown in the middle of the figure.  Here the free energy has two
  minima, an unfolded one and a folded one, and there is no glass
  transition during the folding process. The free energy functions in
  the type II scenarios, shown at the bottom of the figure, also have
  two minima but the protein undergoes a glass transition during the
  folding process.  In a type IIA scenario, shown at the left , the
  glass transition occurs after the thermodynamic folding bottleneck
  at $n_{th}^{\ddagger}$.  In a type IIB, shown at the right ,
  $n_{th}^{\ddagger}>n_g$, making the folding protein glassy before
  the thermodynamic folding bottleneck is reached.}
\label{fig:scenarios}
\end{figure}

How do these considerations of the location of a second order phase
transition corresponding to an ideal glass transition {\em along} the
folding coordinate relative to the extrema of the unimodal and bimodal
free energy functions affect the kinetics of folding?  We see that
there are several distinct folding scenarios, which are illustrated in
Figure~\ref{fig:scenarios} and which we now discuss.  As stated above,
for a unimodal free energy function, downhill folding, the rate of
folding will depend mainly on the lifetimes of the individual
microstates.  We call this situation a Type 0 scenario.  It is
analogous to spinodal crystallization studied in materials
science~\cite{oxtoby}.  In this case the unfolded state is unstable;
from almost any configuration there is a conformational change that
will lower the energy with little cost in entropy.  Never the less,
this type of folding transition can still have a folding bottleneck,
like the folding transitions in a bimodal free energy function if the
diffusion constant becomes small, as in a glass transition.  The
difference here is that the folding bottleneck in a Type 0 transition
will be {\em entirely} kinetic, so $n_{kin}^{\ddagger}$ will occur at
the maximum of $\overline{t}(n)$.  In contrast, for a bimodal free
energy function the folding barrier will have both kinetic {\em and}
thermodynamic contributions.  The Type 0 scenario can further be
broken into two subclasses.  In the first subclass, which we call Type
0A, the glass transition does not occur at any value of $n$.  In this
case the folding is fast and dominated by a single rate, the rate of
going down the free energy gradient.  The kinetics in this regime are
self-averaging.  In the second subclass, which we call Type 0B, the
glass transition occurs before the protein reaches its native state.
Then the first part of the folding is a rapid descent down the free
energy gradient, as before, but the glass transition intervenes and
slows the folding considerably.  The overall kinetics is slower and
multi-exponential because different protein molecules find themselves
stuck in a few different microstates after the glass transition, and
each of these states will fold at a different rate.  Some of the
microstate lifetimes can be very long.  These long-lived microstates
will be observable as kinetic intermediates.  The paucity of occupied
microstates will lead to discrete pathways as shown schematically in
Figure 6.  The kinetic behavior is strongly non-self-averaging, so
mutations easily change the folding kinetics.  Intermediates in one
form of the protein are absent in others.

\begin{figure}
\insfig{.8}{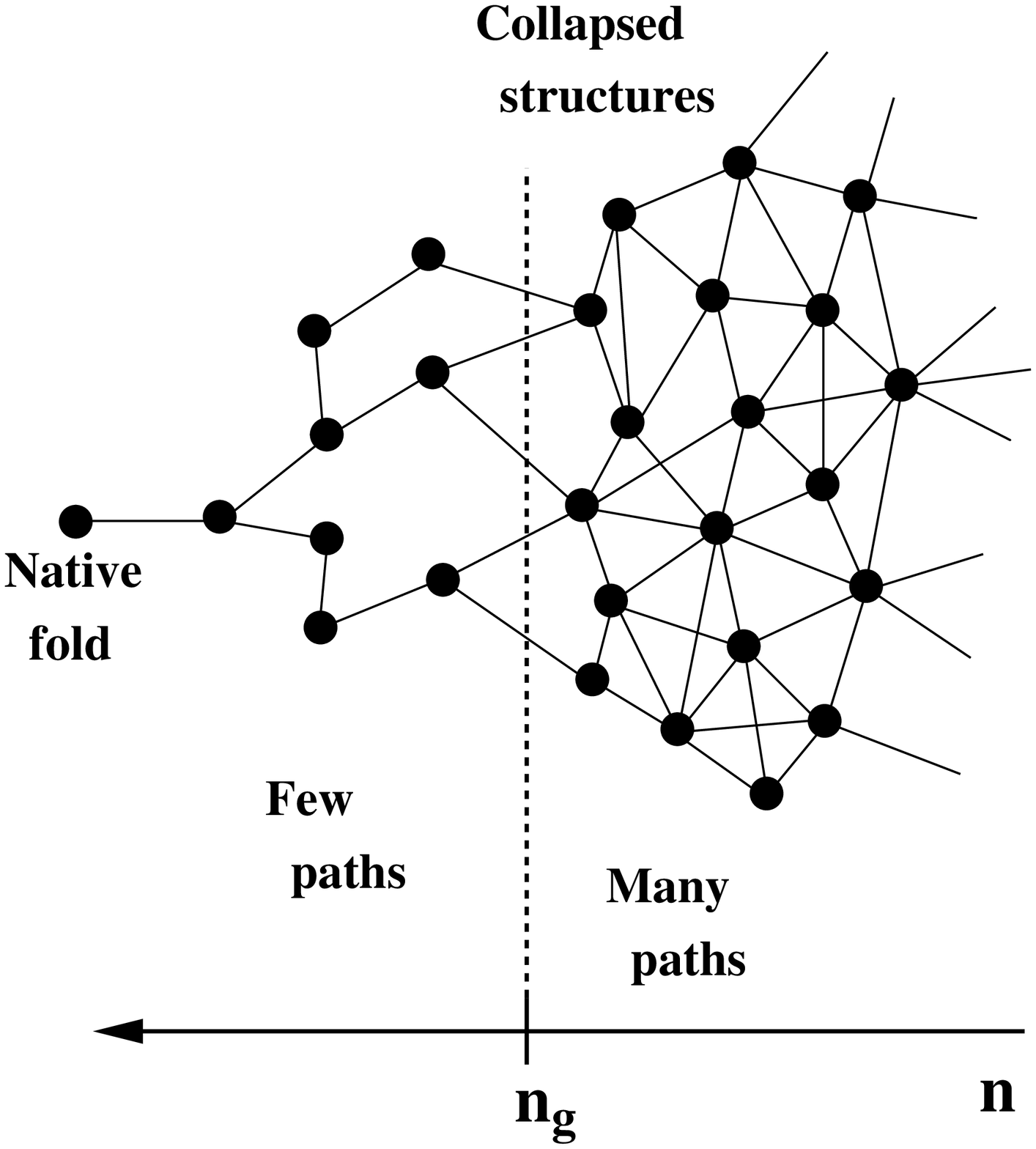}
\caption{A schematic representation of
  the emergence of folding pathways. In this figure the native
  structure is on the left, so that $n$ increases from right to left.
  Before the folding protein reaches the glass transition there are
  many accessible paths between conformations. In this regime each
  molecule would take a different path as it approached the native
  structure.  After the folding protein goes through the glass
  transition it has access to only a few paths, so most molecules will
  take one of a few, or perhaps only one, path to the native
  structure.}
\label{fig:pathways}
\end{figure}

The kinetics of the folding of proteins with bimodal free energy
functions fall broadly into two classes.  In the first of these, which
we call Type I, there is no glass transition at any point in the
folding, just like the type 0A folding scenario noted above.  Type I
scenarios are analogous to nucleation followed by rapid
growth~\cite{oxtoby}. In this case the folding is dominated by a
single rate, the folding time being given by equation
(~\ref{BW_rate}).  The protein has kinetic access to a representative
section of the folding bottleneck, so the rate of folding can be
calculated by considering the rate of folding for a statistical
ensemble of structures at the bottleneck.  In this regime, the protein
can take many possible pathways through the bottleneck, so the overall
folding time will be independent of the initial unfolded configuration
of the protein.  The folding kinetics are self-averaging, so mutations
will have only small effects on folding rates.\footnote{To be more
  precise, rates of individual events depend on the exponentials of
  free energies.  Above the glass transition these free energies
  should all self-average and the significant rates will have a
  log-normal distribution. A few factors of two change in the rate is
  not considered significant here.  In the glassy phase a much wider
  distribution of the logarithm of the rate is anticipated, as pointed
  out by Bryngelson and Wolynes ~\cite{bw_jpc}} In the other class the
glass transition occurs at some point in the folding process.  We call
these folding events Type II.  Type II folding processes are analogous
to nucleation followed by slow growth: a situation much studied in the
metallurgy of alloys~\cite{oxtoby}.  Type II folding scenarios can be
broken into two subclasses, depending on where the glass transition
occurs relative to the thermodynamic bottleneck location
$n_{th}^{\ddagger}$.  Recall that $n_{th}^{\ddagger}$ is the location
of the maximum of the free energy and need not be the same as the
kinetic bottleneck coordinate $n_{kin}^{\ddagger}$ that appears in
equations (~\ref{downhill_rate}) and (~\ref{BW_rate}) for the folding
time.  Thus, for folding at a fixed temperature in a situation where a
glass transition occurs, we expect to find two distinct kinetic
scenarios, one, which we call Type IIA, occurs when $n_{th}^{\ddagger}
< n_g$, and the other, which we call Type IIB, occurs when
$n_{th}^{\ddagger} \geq n_g$.

As the roughness of the energy landscape is increased, a glass
transition occurs between the folding bottleneck and the final folded
state, so that discrete pathways occur after the transition state.  We
call this situation a Type IIA scenario.  In this regime passage
through the folding bottleneck will be dominated by a single rate, but
there may be some non-exponential behavior, and discrete pathways and
kinetic intermediates will be observed in the late stages of folding.

In the Type IIB scenario the protein has already gone through the
glass transition when it reaches the maximum of free energy.  Since
the protein can take only a few pathways after the glass transition,
and these pathways can be different enough to lead to wildly different
folding times, the overall folding time will strongly depend on which
of the few paths to the folded state is taken.  Each of these paths
will have its own kinetic transition state and the free energies of
these states will differ appreciably, {\em i.e.} they will not
self-average.  The importance, and even the meaningfulness, of the
{\em typical} kinetic transition state, $n_{kin}^{\ddagger}$, is
diminished considerably in this regime Therefore we have used the
location of the glass transition relative to $n_{th}^{\ddagger}$
rather than $n_{kin}^{\ddagger}$ in defining the difference between
the Type IIA and Type IIB scenarios.

\section{The Phase Diagram and Protein Folding Scenarios.}

The phase diagram is a powerful tool for understanding protein
folding.  It reduces much of the discussion about folding scenarios in
the previous section to a single, clear, coherent picture which is
useful for thinking about and planning experiments.

\begin{figure}
\insfig{.8}{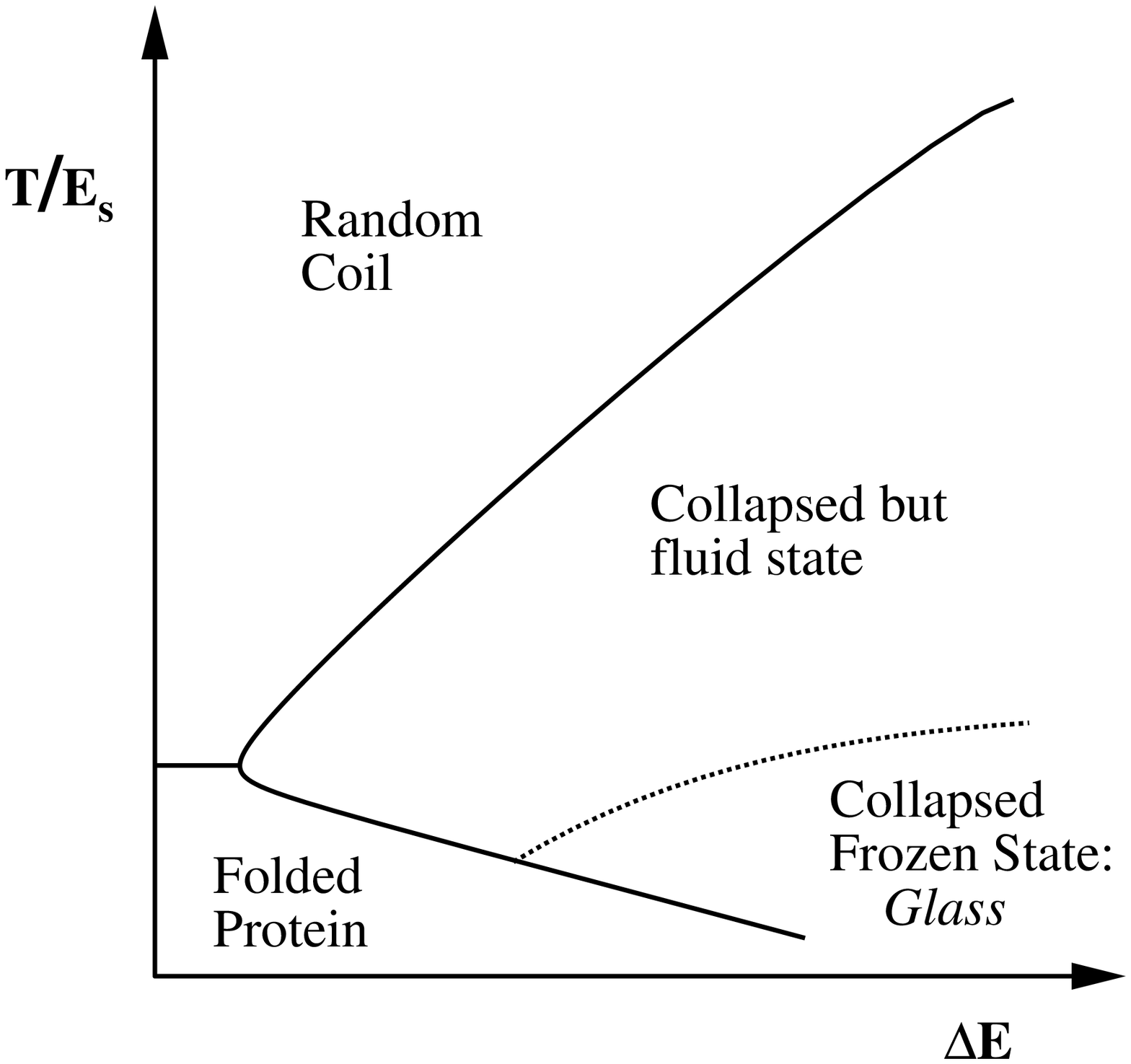}
\caption{Phase diagram for a folding
  protein. The horizontal axis is the energy landscape roughness
  parameter, $\Delta E$, discussed in the text.  The vertical axis is
  the temperature divided by the stability gap $E_s$. The stability
  gap is the energy gap between the set of states with substantial
  structural similarity to the native state and the lowest of the
  states with little structural similarity to the native state.  The
  collapse transition and the (first-order) folding transition are
  represented by solid lines and the (second order) glass transition
  is represented by a dashed line. In comparing this phase diagram
  with experimental phase diagrams, one must bear in mind that both
  $\Delta E$ and $E_s$ are temperature dependent because of the
  hydrophobic force. In addition, the collapse transition depends on
  the average strength of the hydrophobic force, and this is both
  temperature and pressure dependent. The average strength of the
  hydrophobic force could be considered as a third dimension in the
  phase diagram.}
\label{fig:phase}
\end{figure}

The simplified viewpoint of protein folding, using the energy
landscape framework that we discussed in the last section can be used
to classify different mechanisms of protein folding in the laboratory
and in computer simulations.  The analysis discussed above uses only a
single parameter, $n$, to characterize the difference between the
native structure and the unfolded structures.  In fact, native
proteins differ from unfolded ones in several ways, so this requires
the introduction of several different similarity measures in thinking
about folding processes.  It is important, however, that the number of
additional parameters is relatively small, thus giving a reduced
description of the folding process.  Indeed, many of the discussions
of folding pathways have concentrated on these additional similarity
measures or order parameters.  Thus in many pictures of protein
folding, {\em e.g.}, the framework model~\cite{kim_baldwin}, one gives
considerable emphasis to the initial formation of secondary
structures.  In other scenarios, the collapse and formation of
secondary structures are considered to be separate
events~\cite{bw_biopolymers}. Additionally, proteins may consist of
subdomains for which we may discuss the tertiary structure formation
separately.  This is particularly important in hierarchical pictures
of protein folding~\cite{lesk_rose}.  With each of these similarity
measures we can ask the way in which the formation of order is related
to the roughness of the energy landscape and whether the transition
occurs through many pathways or through a small number of distinct
pathways.  It is helpful to consider a phase diagram like the one
illustrated in Figure~\ref{fig:phase}.

In this phase diagram, we plot the possible equilibrium states of a
protein as a function of temperature and roughness of energy
landscape.  The phase diagram contains a region of random coil, a
collapsed phase, folded region with transition lines between these
places, as well as a dotted line indicating the presence of frozen
glassy state.

A given protein will exist at equilibrium somewhere in this phase
diagram, thus the diagram tells us the final state which we would
obtain in an experiment.  The folding process begins by starting in a
configuration characterized by one of the regions on this diagram, but
is carried out at a temperature such that the folded protein is the
lowest free energy state.  The roughness of the energy landscapes is
important in determining the equilibrium phase but plays a bigger role
in the kinetics of the folding process as described before.  In the
lefthand part of the diagram, folding will occur by a Type I mechanism
in which discrete pathways are not observed.  As the roughness is
increased, the folding can occur by a Type IIA mechanism in which
discrete pathways occur after the transition state.  As the roughness
of the energy landscape increases more, and the equilibrium glass
transition occurs before the transition state is reached so the
folding occurs through a Type IIB mechanism in which discrete pathways
are observed and misfolded states play a role in the dynamics.
Structurally unique thermodynamic transition states (bottlenecks) can
occur only if $T<T(n_{th}^{\ddagger})$, {\em i.e.}, if the folding is
Type IIB, because that is the only case where there are order one
accessible paths through the folding bottleneck.  In all other folding
scenarios, there are many accessible paths through the folding
bottleneck, hence many possible transition states.

The temperature at which the folding experiment takes place also plays
an important role in whether a Type 0, Type I or Type II scenario for
folding is observed.  At low temperatures, (relative to the roughness
energy scale) one expects to see nonexponential kinetics
characterizing a Type IIB scenario.  On the other hand, at higher
temperature, at the midpoint of the folding transition, one expects to
see Type I or IIA mechanisms to be more prevalent.  Since ruggedness
only appears when contacts are made, when there is little frustration,
as well as little average driving force towards hydrophobic collapse,
a Type I mechanism is most probable.  This is very close to the
framework model ~\cite{kim_baldwin} or diffusion
collision-picture~\cite{kw_nature,kw_biopolymers,weaver_1,weaver_2,kw_recent}
that was so often thought to describe protein folding.  In the
original versions of such models, only correct structures are formed
initially and these can dock to form completed structures.  Such a
highly unfrustrated situation seems to be uncommon and certainly does
not occur in the computer simulations of protein-like models.

Good folding sequences are ones that have a strong free energy
gradient leading to the ground state structure.  To achieve this they
must separate in energy the native conformation and those
conformations that are structurally similar to the native conformation
from the bulk of most of the other conformations with no structural
similarity to the native conformation.  Goldstein, Luthey-Schulten and
Wolynes have shown that this qualitative criterion is equivalent to
finding sequences that maximize $ T_f / T_g $ for a suitable
simplification of the Bryngelson-Wolynes model~\cite{glsw1}.  Notice
that the energy gap that is being maximized when $ T_f / T_g $ is
maximized is {\em not} the energy gap between any two specific states,
but rather the gap between the {\em set} of states with substantial
structural similarity to the native state and the lowest of the {\em
  set} of states with little structural similarity to the native
state.  We call this gap the ``stability gap'' ($E_s$).  The stability
gap should not be confused with the energy gap between the native
configuration and the configuration with the next highest energy.
This state will usually be native-like itself.  There are too many
fluctuations in the folded state for this two-configuration energy gap
to have any significance for protein folding or
stability~\cite{mccammon_harvey,dave_2,hans_f,karplus_mccammon}. In
fact, Frauenfelder and collaborators have interpreted the results of
their experiments on {\em folded} proteins in terms of a hierarchy of
``substates''.  These substates correspond to slightly different
structures found in the population of folded proteins~\cite{hans_f}.
Evidence for the highest level of this hierarchy has been seen in
protein folding simulations~\cite{dave_3}.  Unfortunately, this issue
of energy gaps has been clouded by lattice simulations that have
studied the energy spectrum of only the maximally compact
states~\cite{nature_paper,ssk_jmb}. Since the maximally
compact states are a small fraction of all possible states and since
the they are often not dynamically connected, (in the sense described
in Section 2) ~\cite{dave_4,leopold} the interpretation of the results
of these simulations requires more subtlety than has been found in the
literature so far.  Notice, however, that since local excitations from
a maximally compact state are not themselves maximally compact, the
energy gap between the native state and the next lowest energy
maximally compact state is often correlated with the stability gap.
Thus, the results of these simulations can be interpreted as a
confirmation of the older and more general idea the sequences with
large stability gaps fold quickly at the equilibrium folding
temperature~\cite{bw_pnas,bw_jpc,bw_biopolymers,glsw1,glsw2}.

In the Bryngelson-Wolynes energy landscape the stability gap is a
tautological consequence of the greater degree of stability of
native-like interactions demanded by the principle of minimal
frustration.  Goldstein, Luthey-Schulten and Wolynes calculated a set
of parameters that maximized $ T_f / T_g $ for the model used in their
protein structure prediction algorithm, and found that these
parameters gave excellent results for practical structure prediction,
in accord with the predictions of the theory.  In addition, molecular
dynamics calculations using associative memory Hamiltonians optimized
in this way reliably gave native-like structures~\cite{glsw1,glsw2}.
These results provide independent evidence that sequences that satisfy
this criterion (of having a large stability gap) should be good
folding sequences.  This work also is a good illustration of the power
of using energy landscape ideas to help solve practical protein
folding problems.  We also mention that the stability gap idea has
been used by by Wodak and co-workers to predict persistent secondary
structures in small peptides relevant to early folding
events~\cite{wodak}.

The phase diagram, of course, becomes more complex as additional order
parameters or similarity measures are used to characterize the folded
states.  The phase diagram is a useful way of thinking about any
folding process because it allows us to consider the couplings between
the various order parameters as well.  For instance, as one sees in
the computer simulations one can first have a collapse which is
ascribed by a single-order parameter radius of gyration, followed
later from this collapsed phase by a transition to a unique folded
protein structure~\cite{dave_4,sk_proteins1,sk_proteins2,socci}.  The
coupling between these two parameters is crucial in obtaining that
sort of description. The so-called molten globule intermediates which
are often an ensemble of individual configurations really should be
described by these additional order parameters~\cite{ptitsyn}.

\section{Energy Landscape Analysis of Folding Simulations}

Simulations of simple protein-like lattice models provide an ideal
ground to illustrate the energy landscape ideas.  Lattice models have
a venerable
history~\cite{lau_dill_macro,lau_dill_pnas,chan_dill_sss,go1,go2,go3,go4,mj,ksy,sk_science,kms,sk_jmb,rs}.
There is widespread agreement that they capture some of the underlying
physics of protein folding.  There are also excellent reviews that
discuss lattice simulations in the context of the general problem of
understanding protein
folding~\cite{go_annrev_biophys,skolnick_annrev_pchem,chan_annrev_biophys}.
Many groups have interpreted their simulation results using some of
the qualitative and semi-quantitative ideas of energy landscape
analysis, finding features in agreement with the overall picture that
we have just
discussed~\cite{dave_4,dave_2,dave_3,chan_dill_tst,nature_paper,dave_5,chan_dill_homo,covell}.
Here we illustrate this kind of discussion by focusing on some recent
results of Socci and Onuchic which find evidence for specific features
arising from energy landscape analysis~\cite{socci}.  In addition,
these simulations provide an excellent example of the kind of
quantitative analysis which should be carried out for real
experimental data.  We will use simplified quantitative relations that
can be deduced from the energy landscape analysis.  This sort of
quantitative analysis should also be carried out for laboratory
experiments, but in the laboratory the temperature dependence of the
various free energy contributions must also be included explicitly for
a fully convincing analysis.  Simulations based on reduced models
avoid these issues since the energy function is itself not temperature
dependent.

The simulations were performed on polymers that were $27$ monomers
long which have maximally compact states of $3\!\times\!3\!\times\!3$
cubes.  Because the configurations on the $3\!\times\!3\!\times\!3$
cube can be completely enumerated in a reasonable amount of computer
time, the energy landscape among the maximally compact states can be
explored in great detail.  This $27$ monomer cubic simulation has been
a paradigm of study in this field because of this
feature~\cite{sg_nature,sg_jcp,nature_paper,ssk_jmb,eugene_prl}.
The simulations of Socci and Onuchic contain two monomer types.  Pairs
of monomers that were nearest-neighbors on the lattice but not
connected along the chain contributed an interaction energy to the
potential.  The potential for the two monomer code was is $-3$ for
contacts between monomers of the same type and $-1$ for contacts
between different types.  The folded configuration was taken to be
maximally compact configuration with the lowest energy.

\begin{figure}
\insfig{.8}{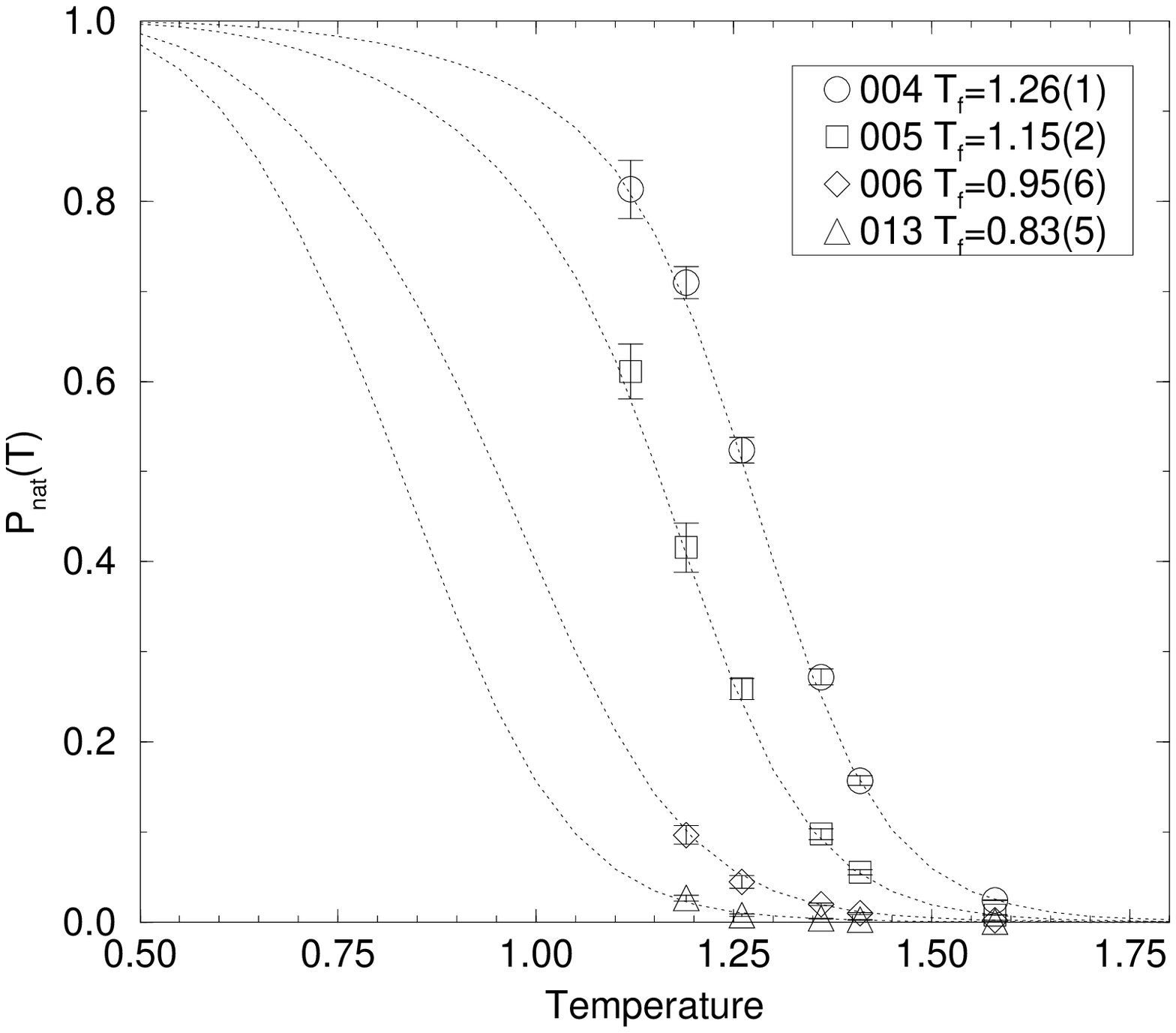}
\caption{Folding curves for four of the sequences used in the
  simulation. The probability of a sequence occupying the native
  structure is plotted on the vertical axis versus temperature on the
  horizontal axis.  The folding temperature, $T_f$, is defined as the
  temperature where $P_{nat}(T_f) = 0.5$, i.e., the probability of the
  occupancy of the native structure is one-half.  The numbers in
  parenthesis indicate the uncertainty in the last digit.}
\label{fig:thermo}
\end{figure}

The characteristic energy scales and temperatures for different
sequences is easily obtained for these models.  The folding
temperature, $T_f$, may be defined in the usual way as the temperature
at which population in the folded configuration is equal to the
populations in all other configurations.  These populations can be
obtained by a Monte Carlo sampling procedure for each of the
sequences.  The folding temperatures correlate rather well with the
energy of the folded configuration.  This is shown in
Table~\ref{tab:seq}.  Figure~\ref{fig:thermo} shows the equilibrium
folding curves for these sequences.

\begin{table}
\footnotesize
\begin{tabular}{lcrrrr}
  Run & Sequence & $E_{\rm min}$ & {${\tau_{\!\rm min}}$\ \ \ } &
  $T_g$ &
  {$T_f$\ \ \ } \\ \hline
  002 & ABABBBBBABBABABAAABBAAAAAAB & -84 & $2.0\times10^7$ & 1.00 &
  1.285(15) \\
  004 & AABAABAABBABAAABABBABABABBB & -84 & $1.6\times10^7$ & 0.96 &
  1.26(1) \\
  005 & AABAABAABBABBAABABBABABABBB & -82 & $2.3\times10^7$ & 0.98 &
  1.15(2) \\
  006 & AABABBABAABBABAAAABABAABBBB & -80 & $5.2\times10^7$ & 1.07 &
  0.95(6)\\
  007 & ABBABBABABABAABABABABBBABAA & -80 & $9.3\times10^7$ & 1.09 &
  0.93(5)\\
  013 & ABBBABBABAABBBAAABBABAABABA & -76 & $9.7\times10^7$ & 1.01 &
  0.83(5)
\end{tabular}
\caption{\protect\footnotesize The various sequences used in
  this paper. The last four (005, 006, 007, 013) were generated at
  random. Sequence 002 is from reference 124.  Sequence 004 is a
  single monomer mutation of 005 ($B_{13}\rightarrow A$). Both 002 and
  004 have the lowest energies possible for the potential used and
  have native states that are completely unfrustrated, i.e. every
  native contact is individually stabilizing. ${\tau_{\!\rm min}}$ is
  the fastest folding time for each sequence. $T_g$ is the glass
  transition temperature (calculated with a ${\tau_{\!\rm
      max}}=1.08\times10^9$).  $T_f$ is the folding temperature
  calculated using the Monte Carlo histogram method. The numbers in
  parenthesis indicate the uncertainty of the last digit.}
\label{tab:seq}
\end{table}

A kinetic glass transition temperature can be defined without
appealing explicitly to the energy landscape analysis.  Just as in a
laboratory, a kinetic glass transition temperature is defined by
asking where a characteristic timescale in the problem exceeds some
large value.  In the simulations the maximum running time was
$\tau_{\rm max} = 1.08\times10^9$ Monte Carlo steps.  This number was
chosen because it was significantly longer than the folding times over
a broad range of temperatures. It would be appropriate to define the
characteristic time through the typical time for a large-scale
rearrangement.  However, it is simpler here to use the folding time
itself as a time-scale.  A kinetic glass transition temperature,
$T_g$, then is defined by the criterion $\tau_f (T_g)$ is $(\tau_{\rm
  max}+\tau_{\rm min})/2$ where $\tau_f (T)$ is the folding time at
temperature $T$.  As you can see from Table 1, this transition is
nearly self-averaging, that is, it depends very little on the
particular sequence which is studied, and is roughly $1.0$.

According to the energy landscape analysis this kinetic glass
transition is most strongly influenced by the thermodynamic glass
transition.  The simulations bear out this expectation.  Changing the
fiducial cut-off time by a factor of $8$ causes only a $10$ percent
change in the kinetic $T_g$.  Similarly, small changes to the
algorithm for selecting the moves have a small effect on
$T_g$.\footnote{Only if the number of crankshaft moves is reduced to
  less than $10$ percent of the corner moves is there any very
  dramatic change in $T_g$.}

The thermodynamic glass transition of the BW analysis depends on the
entropy and roughness energy scale of the compact states.  This
thermodynamic $T_g$ is also a self-averaging quantity.  Using only the
maximally compact cube states, one obtains $T_g \approx 1.17$.  This
estimate of $T_g$ is likely an upper bound, since semi-compact states
also contribute to the entropy.  At the same time, kinetic constraints
could create additional restrictions on this connectivity.  These
effects seem to cancel, so the kinetic and thermodynamic glass
transitions are rather close and one can take them both to be
approximately $1$ in analyzing the figures.

\begin{figure}
\insfig{1}{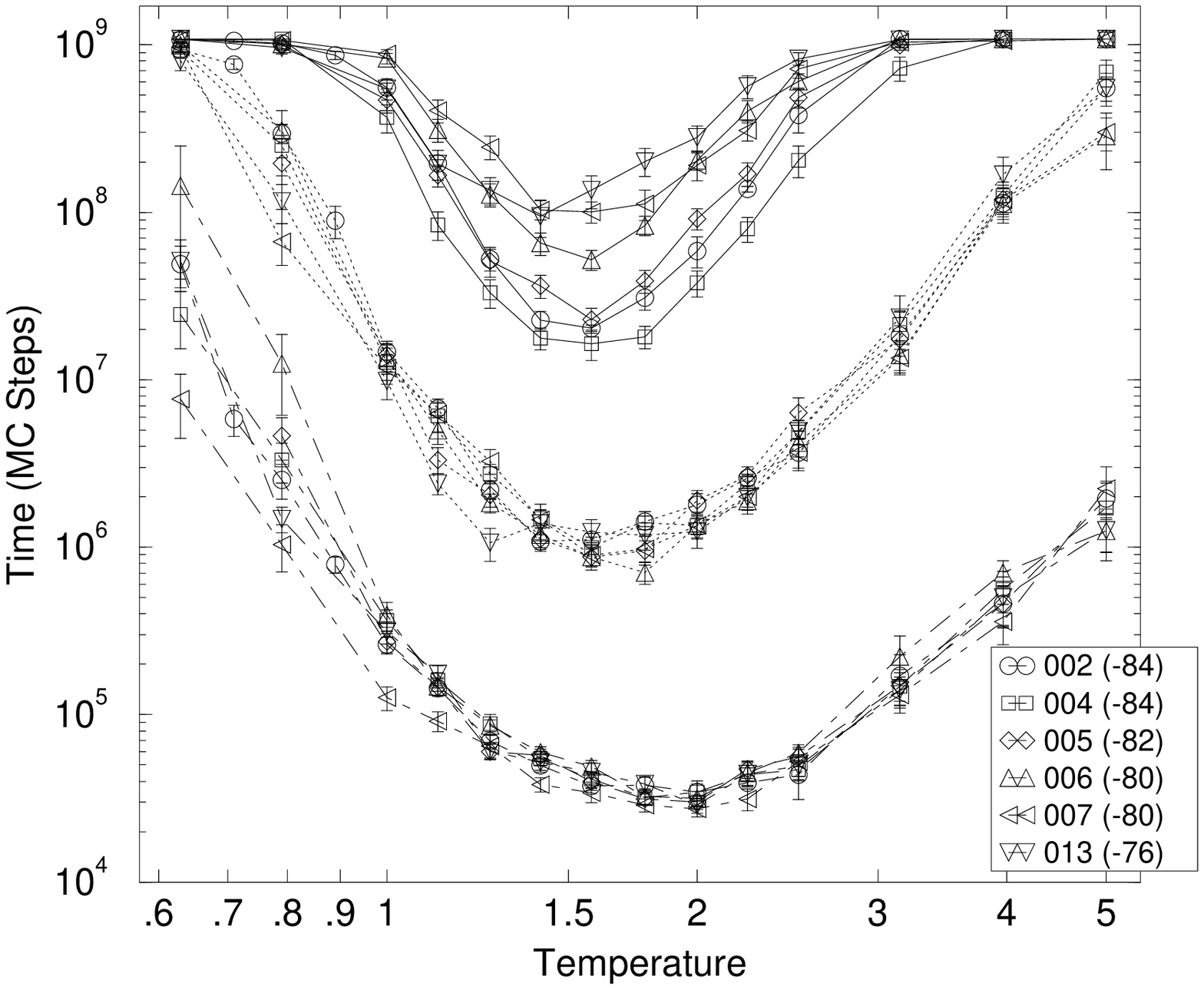}
\caption{
  Plots of important times (in Monte Carlo steps) against temperature
  for the sequences used in our simulations.  The top curves are the
  folding times $\tau_f$ (the number of steps required to reach the
  native structure for the first time).  The saturation at the wings
  of the curves occurs because runs were stopped at a maximum time of
  $1.08 \times 10^9$ Monte Carlos steps. The other curves are plots of
  collapse times.  The middle curves are the times required for the
  sequences to reach a conformation with 28 contacts for the first
  time. Similarly, the bottom curve is the time required to reach a
  conformation with 25 contacts for the first time. Notice that there
  is a much greater time spread in the folding curves than in the two
  collapse curves.}
\label{fig:times}
\end{figure}

Figure~\ref{fig:times} shows a plot of the folding time, that is, the
time that takes for a random unfolded initial condition to reach the
native structure, for different temperatures.\footnote{Technically,
  these times are mean-first-passage-times.} Since the $27$ monomer
length heteropolymer is so small, it is possible to analyze folding
both above and below $T_f$ for quite a range of temperatures.  Above
$T_g$ the folding process is essentially an uphill one but with a
modest slope.  The first noticeable feature about the folding data is
that they are strongly sequence dependent at intermediate
temperatures.  In the simulations, and folding times greater than a
maximum of $\tau_{\rm max}$ were assigned the folding time $\tau_{\rm
  max}$.  This is the origin of the saturation at the high and low
temperature ends of these curves.  At high temperature the folding is
slow because it is so strongly uphill entropically.  At low
temperatures the folding is slow because of the roughness of the
energy landscapes for all of these sequences.  Another characteristic
feature, however, is that the folding time at intermediate
temperatures is most strongly correlated with the stability of the
folded state for each of the sequences.  The fastest folding sequence
has the highest folding temperature, while the slowest has the lowest
folding temperature.  Indeed, the slowest folding sequence has a
folding temperature less than the glass transition temperature.

Also plotted in Figure~\ref{fig:times} are two different collapse
times for the same sequences.  The lower curve is the time that it
takes the sequence to encounter, for the first time, a structure with
$25$ contacts.  The middle curve is the time needed by the protein to
achieve any maximally compact $28$ contact cube.  The remarkable
qualitative feature of these collapse time curves is that at the
moderate to high temperatures where the folding times vary greatly,
all of the sequences have essentially the same collapse times.  In
this temperature range collapse is a self-averaging process that
depends primarily on the average composition of the protein molecules.
Another remarkable feature, however, is that the collapse time begin
to fluctuate greatly between different sequences at and below the
kinetic glass transition temperature.  The energy landscape analysis
suggests that individual transition times between states fluctuate
greatly below $T_g$, and this is reflected in the collapse process.
The distribution of folding times becomes broader as you approach
$T_g$, reflecting the emergence of a multi-exponential collapse
process.  We note that Flanagan {\em et al.} have observed sequence
dependent collapse in staphylococcal nuclease~\cite{flanagan}.  This
suggests the phase observed is near its glass transition.

A rough quantitative understanding of these data for folding and
collapse comes from energy landscape analysis.  The availability of
both folding and collapse results allows us to roughly separate
features connected with the glassy dynamics from the thermodynamic
changes that also result from rough energy landscapes.  The first
important observation is that both folding and collapse times give
parabolic Arrhenius plots, just as most experimental data do for the
forward and reverse rate of folding~\cite{creighton_pain_book}.  In
the laboratory this curvature is usually ascribed to the thermodynamic
dependence of the effective interactions, the difference of heat
capacity between the folded and unfolded states arising from the
hydrophobic effect.  Since the force laws in the simulation are taken
to be {\em independent} of temperature, the temperature dependence of
the hydrophobic effect is not at all involved in the simulation data.
The simulation of Miller {\em et al.} also effectively finds a curved
Arrhenius plot~\cite{miller}.  A simple analysis can be carried out by
assuming that the location of the folding bottleneck,
$n_{kin}^{\ddagger}$, is independent of temperature.  Roughly speaking
then, the folding time will be given by equation (\ref{BW_rate}) with
the energy barrier $F_{kin}^{\ddagger}$ given by the difference in
free energy between the folding bottleneck states and the free energy
of the bottom of the unfolded free energy minimum, {\em i.e.}, the
lowest free energy unfolded states.  This involves motion on the free
energy gradient for the reaction coordinate based on the number of
correct contacts.  At this level of analysis, the collapse time can be
treated in a similar way using the total number of contacts of any
kind as a reaction coordinate.  In the temperature range of $1.0$ to
$2.25$ (the reason for considering this temperature range will become
clear below) the time required for collapse to configurations with 25
contacts varys by a factor of less than $4$, indicating that there is
little, if any, free energy barrier to collapse.  Therefore, collapse
is essentially downhill in free energy and behaves like a Type 0A
scenario.  The dynamical reorganization timescale will become longer
as the protein becomes more compact because excluded volume has a
stronger effect on dynamics in compact states.  Therefore, in the
generalized transition state approximation of Section 4, the collapse
time will be given by equation (\ref{downhill_rate}) for the time for
a downhill process,
\begin{equation}
  \tau_{collapse} = \overline{t}_{collapse},
\end{equation}
where $\overline{t}_{collapse}$ is the typical lifetime of an
individual microstate in a random collapsed state.  For the purposes
of calculating the barrier height, $F_{kin}^{\ddagger}$, we set the
free energy of the bottom of the unfolded free energy minimum equal to
the free energy of the collapsed states.  Then the folding time
involves the free energy difference of the folded and compact
configurations.  Another way of obtaining a folding time that depends
on this free energy difference is to consider folding to be a three
state unimolecular reaction, $random \: coil
\rightleftharpoons collapsed \rightarrow
folded$, where the second step, $collapsed \rightarrow folded$ is
rate-limiting.  The data is consistent with such a reaction scheme.

\begin{table}
\footnotesize
\begin{tabular}{lccc}
  & $A$ & $B$ & $C$\\
  Run & $= S_0(n_{kin}^{\ddagger}) - S_{0,collapse}$
  & $=\overline{E}(n_{kin}^{\ddagger})-\overline{E}_{collapse} $
  & $=(1/2)[\Delta E(n_{kin}^{\ddagger})^2-\Delta E_{collapse}^2]$
  \\ \hline
002 & 19.0 & -34.9 & 23.8 \\
004 & 17.0 & -30.1 & 20.1 \\
006 & 16.9 & -25.2 & 16.1 \\
007 & 16.1 & -22.2 & 15.0
\end{tabular}
\caption{The coefficients of the parabolic fits,
  $ \log(\tau/\tau_{collapse}) = A + B/T + C/T^2 $, to the data shown
  in Figure~\protect\ref{fig:times}.  The sequence numbers refer to
  the sequences displayed in Table 1.  The column headings also show
  the physical chemical interpretations of the coefficients given in
  equation (16) in the text.}
\label{tab:cof}
\end{table}

\begin{figure}
\insfig{1}{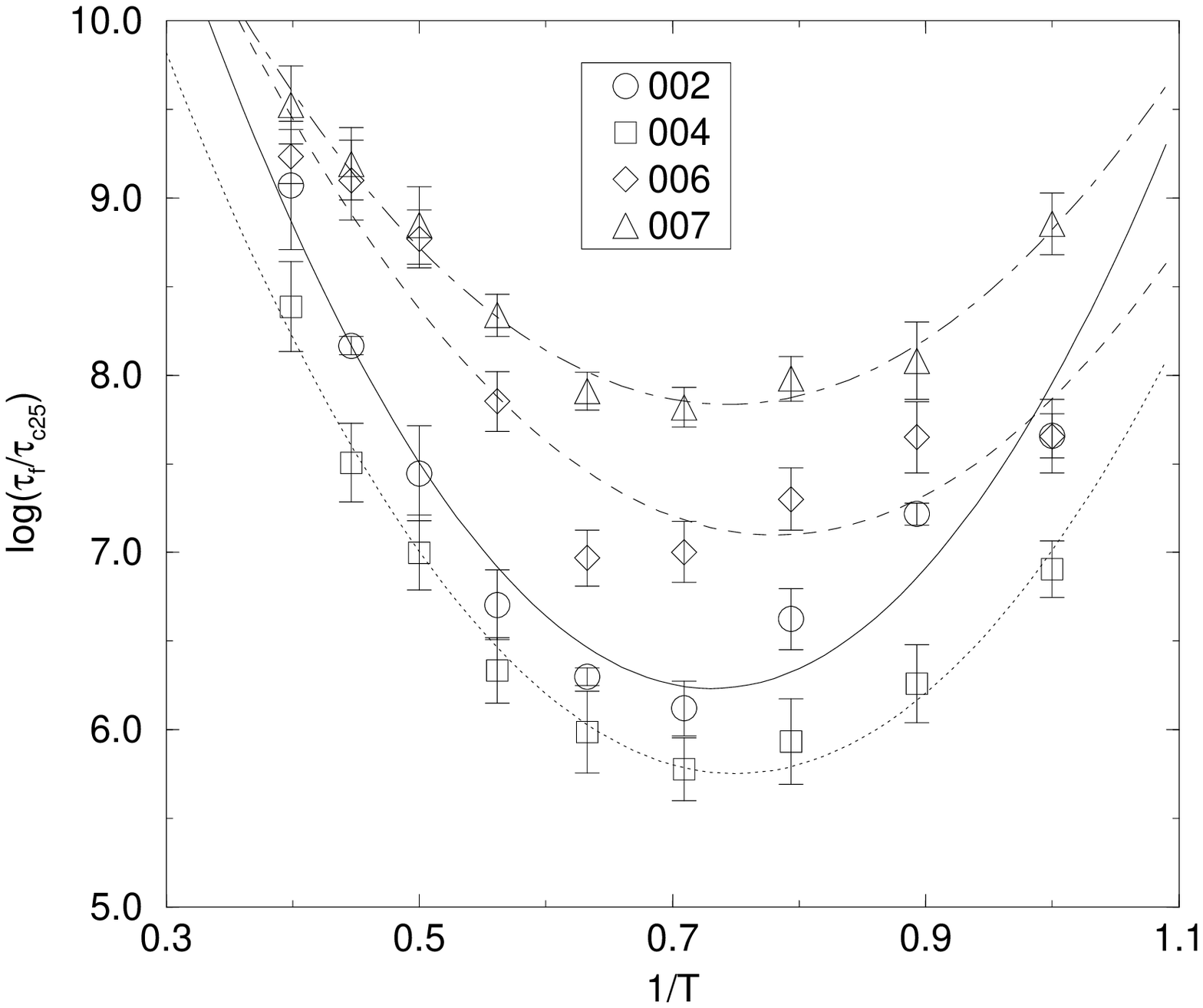}
\caption{
  The logarithm of the ratio of the folding time to the time for
  collapse to 25 contacts against the inverse temperature.  The lines
  are parabola fits to the data. The coefficients of these parabolas
  are shown in Table~2.}
\label{fig:glass}
\end{figure}

We can eliminate the purely dynamical factors by taking the ratio of
the folding to the collapse time and assuming that
$\overline{t}(n_{kin}^{\ddagger}) \approx \overline{t}_{collapse}$.
Then using the equation (\ref{BW_rate}) for the folding and collapse
times and using equation (\ref{free_energy}) for the free energy
predicts that a plot of the logarithm of the ratio of the folding to
the collapse times to be parabolic,
\begin{eqnarray}
  \log \left(\frac{\tau}{\tau_{collapse}}\right) =
  &-&[S_0(n_{kin}^{\ddagger}) - S_{0,collapse}] \nonumber \\
  &+&\frac{[\overline{E}(n_{kin}^{\ddagger})-\overline{E}_{collapse}]}
  {k_B T} \nonumber \\ &-&\frac{[\Delta E(n_{kin}^{\ddagger})^2-\Delta
    E_{collapse}^2]} {2(k_B T)^2},
\label{parabola}
\end{eqnarray}
where the subscript $collapse$ indicates that the quantity is
evaluated in a random collapsed state.  The log of this ratio is
plotted {\em versus} $1/T$ in Figure~\ref{fig:glass}.  We show here
the data only between temperatures 1.0 and 2.25 because outside this
range the folding times exceed the time used as a cut-off in the
simulations.  These curves can be fit very adequately with parabolas.
The coefficients of the parabolas are shown in Table 2.  In the fit
all of the constant terms are positive and all of the linear (in
$1/T$) terms are negative, which imply the inequalities
$S_0(n_{kin}^{\ddagger}) < S_{0,collapse}$ and
$\overline{E}(n_{kin}^{\ddagger}) < \overline{E}_{collapse}$.  Both of
these inequalities are consistent with the bottleneck for folding
occurring after the collapse, in agreement with both intuition and the
simulation data.  The curvature reflects the value of the roughness of
the energy landscape of the collapsed configurations.  This analysis
shows that for a rough energy landscape, the heat capacity of the
collapsed configurations arises from fluctuations in structure and
corresponding energy differences between collapsed configurations.
The linear term in $1/T$ reflects primarily the enthalpic part of the
activation free energy for achieving a transition state.  It should be
strongly correlated with the stability gap.

One can also check the theory by using independently derived
information about the simulation model to make order-of-magnitude
estimates of the sizes of the coefficients in the parabola fits.  The
constant term is the difference of the configurational entropies of
the collapsed states and the folding transition bottleneck states.
The number of states with 25 contacts has been estimated to be
$10^{9}$, yielding a configurational entropy of $9 \log 10 \approx
21$.  The configurational entropy of the folding bottleneck states is
more difficult to estimate, but it is clearly less than that of the
collapsed states.  Therefore, the constant coefficient is expected to
be of order $10$, {i.e.}, between $\approx 3$ and $\approx 30$.  This
expectation is very well confirmed by Table 2, where the constant
coefficients are see to lie between $16$ and $19$.  The coefficient of
the $1/T$ term is the difference between the average energy of the
folding bottleneck states and the collapsed states.  We have defined a
collapsed state to be a state with $25$ contact and the average
contact energy in our model is $-2$, therefore, the average energy of
a collapsed state is $-50$.  The average energy of a folding
bottleneck state must be greater than the energy of the native state,
which is $-84$ for sequences $002$ and $004$ and $-80$ for sequences
$006$ and $007$.  (See Table 1).  Therefore, we expect the coefficient
of the $1/T$ term to lie between $0$ and $-34$ for the first two
sequences and to lie between $0$ and $-30$ for the later two
sequences.  Table 2 shows the coefficients to lie within these bounds,
within reasonable error estimates.  The coefficient of the $1/T^2$
term is one-half times the difference between the roughnesses of the
the collapsed states and the bottleneck states.  Each interaction
energy in the model differs from the average interaction energy by
$+1$ or $-1$, so the roughness of the set of random collapsed states
with $25$ contacts is $\Delta E = 25$.  The roughness of the
bottleneck states is smaller than this number, but difficult to
estimate.  Thus, we expect the coefficient of the $1/T^2$ term to be
somewhat less than $12.5$.  The values for this coefficient range from
$14$ to $24$, as shown in Table 2.  This estimate is not as good as
the previous ones, but it does give the right sign and
order-of-magnitude, which is the best that can be expected from such
an approximate theory and such simple estimates.

A quantitative relationship between protein folding kinetics and the
thermodynamic stability of the native state can be obtained with
linear free energy
relationships~\cite{moore_pearson,brr,leffler_grunwald,wells}. In the
past these relations have been applied to the interpretation of data
from site-directed mutagenesis
experiments~\cite{matthews_meth_1,fersht_meth,matthews_meth_2,msf}.
They are also the mainstay of the analysis of many other biochemical
reactions~\cite{schellman_bp,szabo,eaton}. In this analysis the
differences in the free energies of the transition states, folded
states and unfolded states for two different sequences are obey the
linear relation
\begin{equation}
  \delta F(n_{kin}^{\ddagger}) = \alpha \delta F(1) - (1-\alpha)
  \delta F(0).
\label{linear_free_energy}
\end{equation}

\begin{figure}
\centerline{\epsfxsize=.4\figurewidth\epsfbox{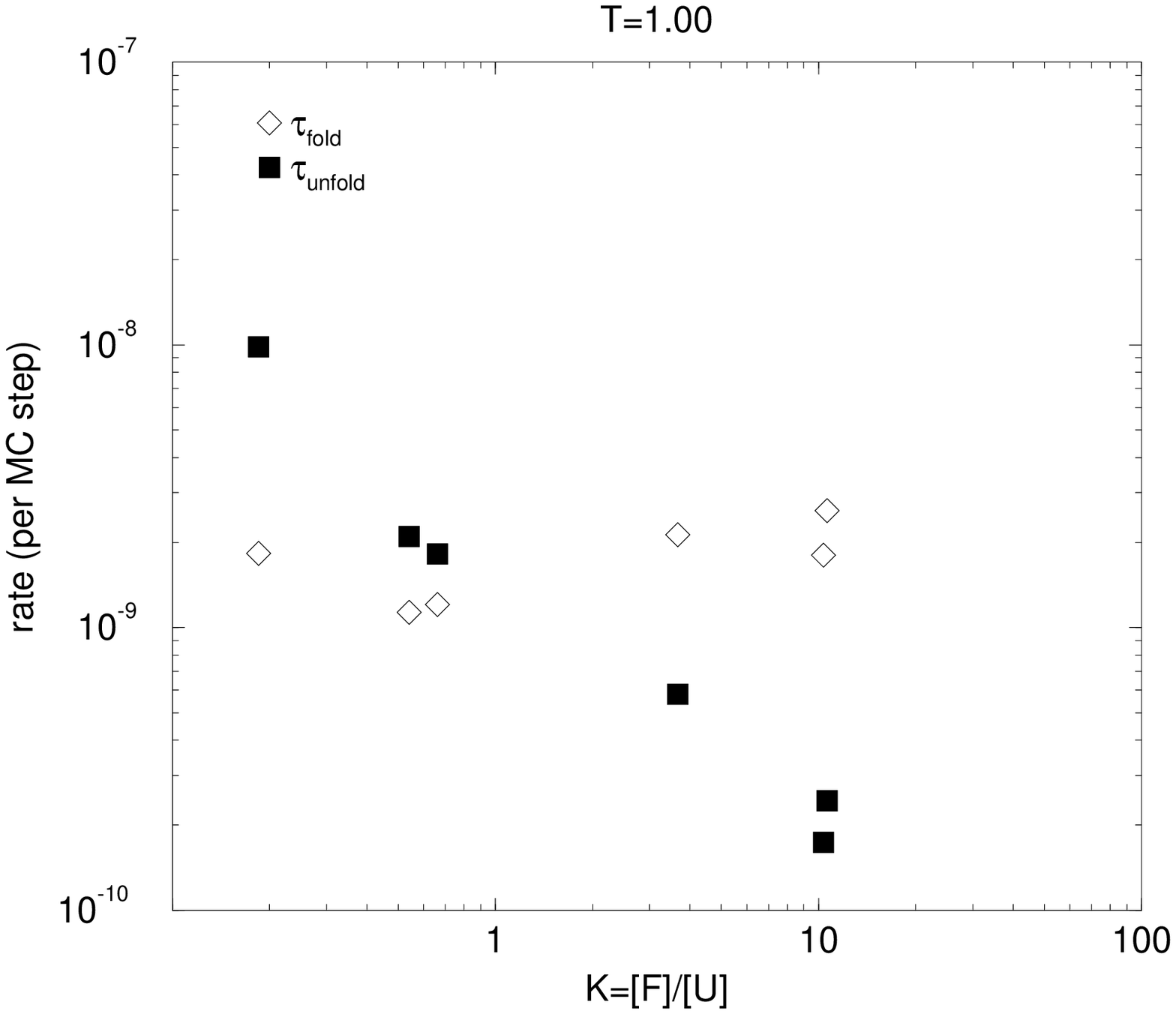}%
            \epsfxsize=.4\figurewidth\epsfbox{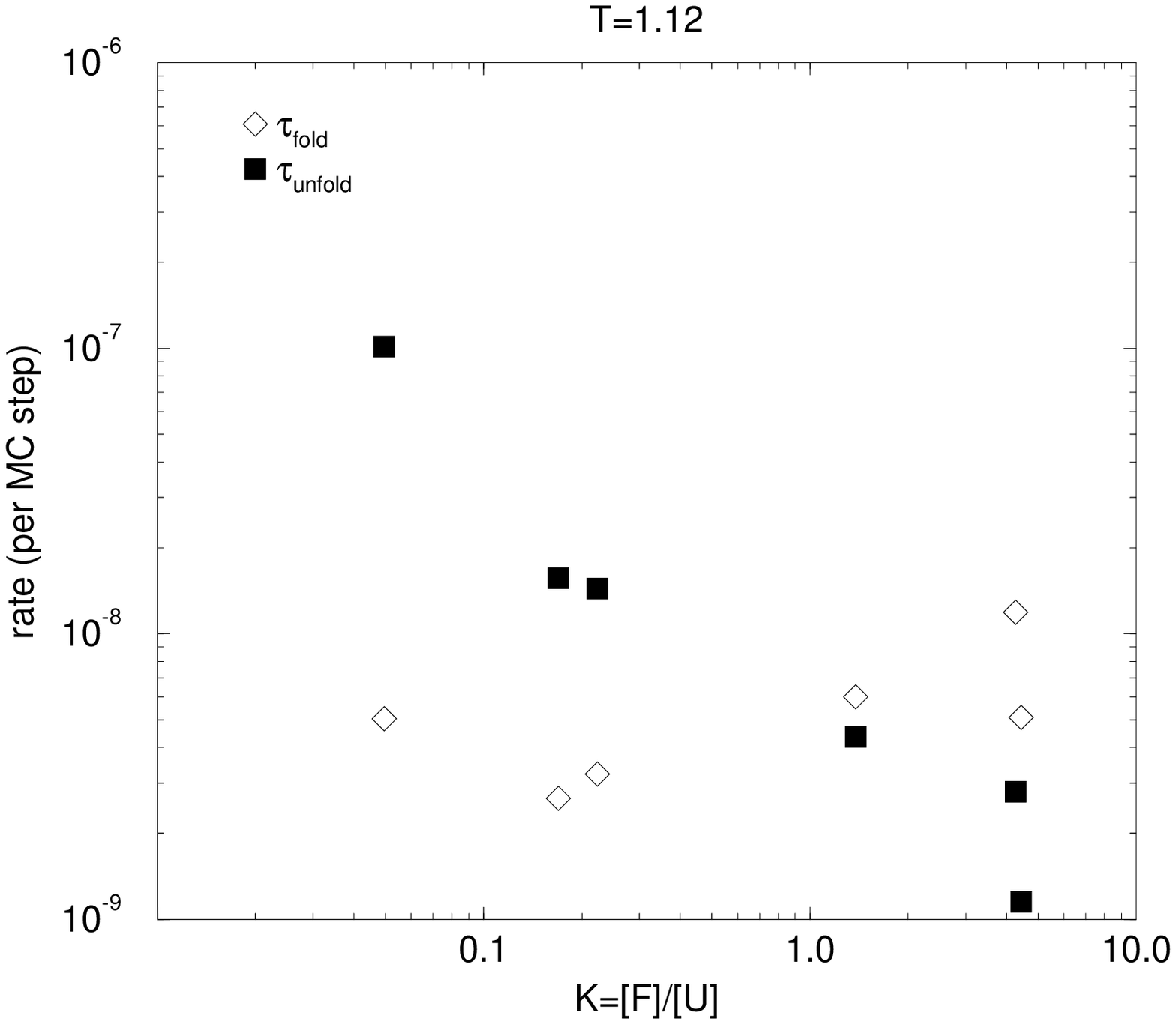}}
\centerline{\epsfxsize=.4\figurewidth\epsfbox{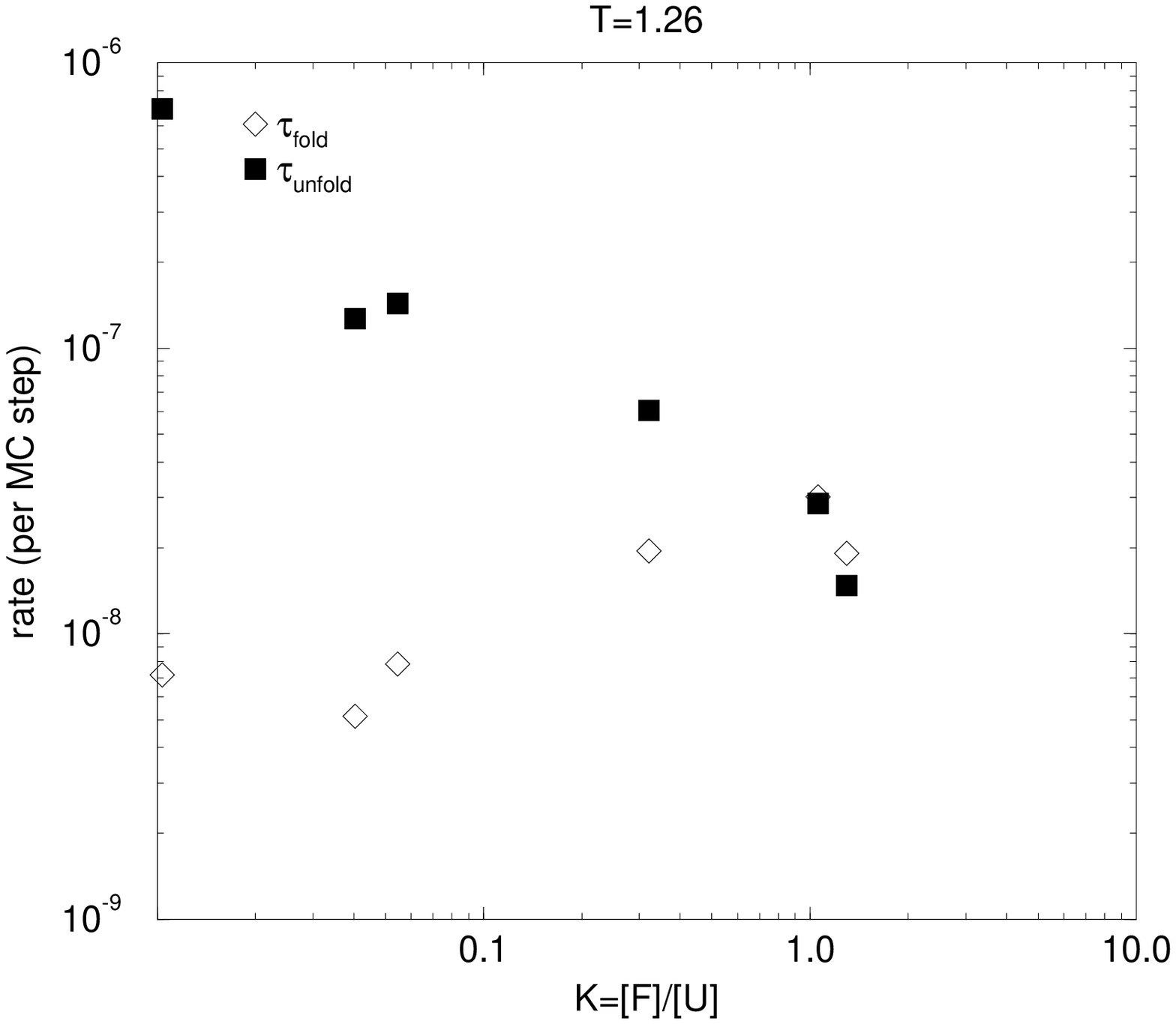}%
            \epsfxsize=.4\figurewidth\epsfbox{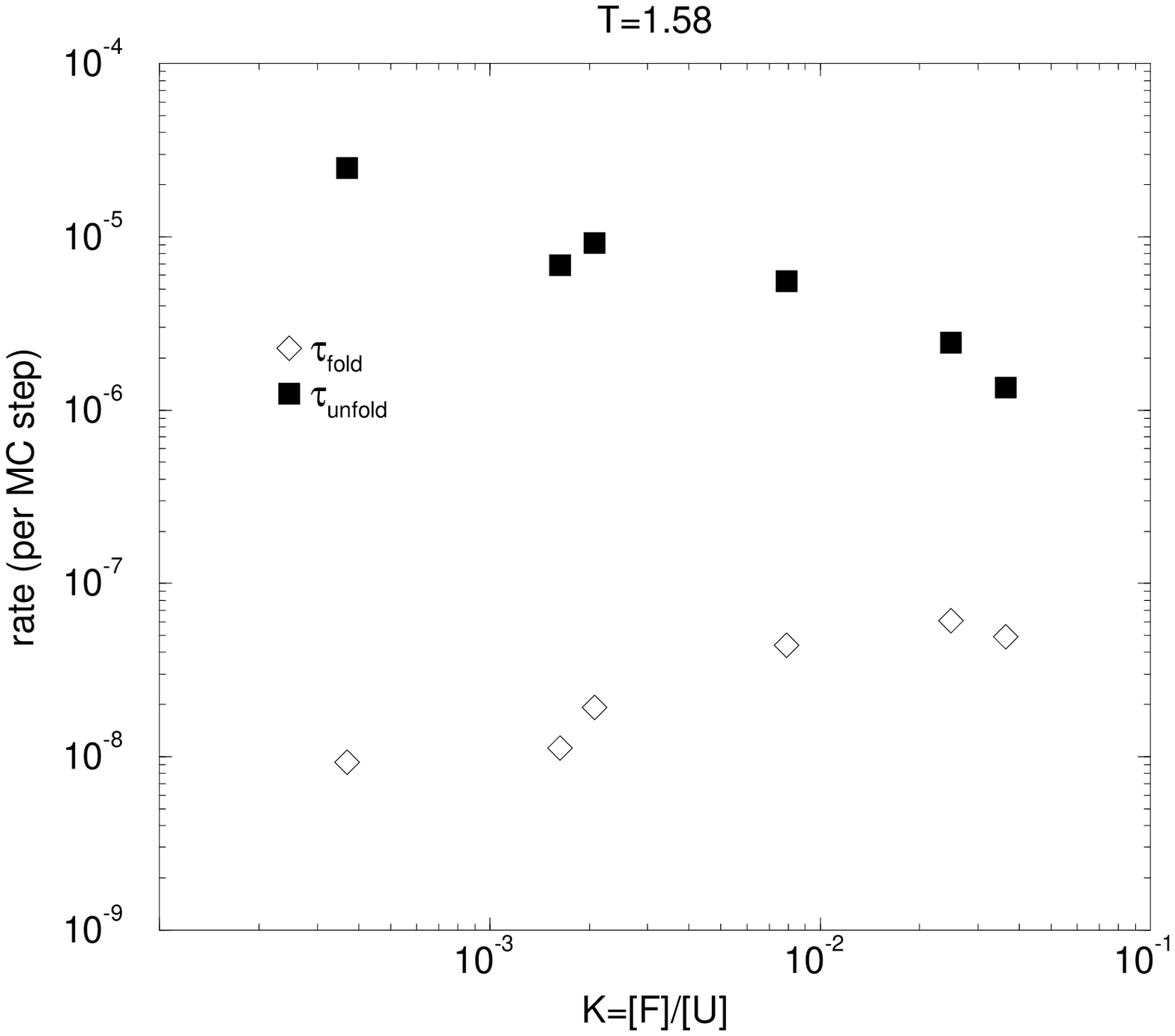}}
\centerline{\epsfxsize=.4\figurewidth\epsfbox{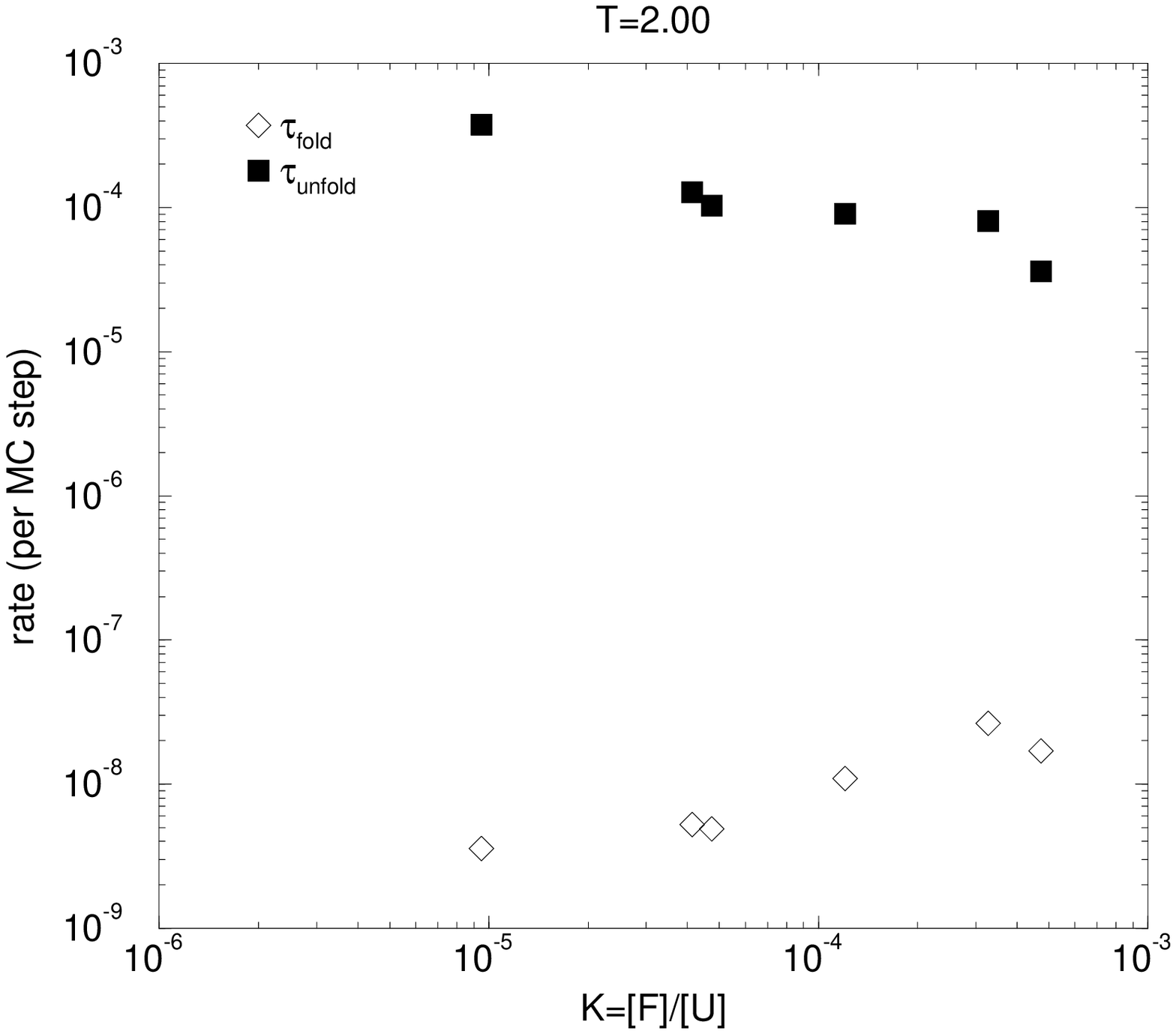}}
\caption{
  Plots for the linear free energy relationship analysis. Each plot
  shows the folding rate against the folding equilibrium constant for
  each of the six sequences studied here.  On the horizontal axis
  $[F]$ represents the probability of the native structure being
  occupied and $[U]$ represents the probability of a non-native
  structure being occupied. Figure 11a is such a linear free energy
  plot for temperature $T=1.0$, that is, at about the glass transition
  temperature for these sequences. The rest of the plots are for
  temperatures above the glass transition temperatures.}
\label{fig:marcus}
\end{figure}

\noindent The transfer coefficient $\alpha$ is a measure of the
resemblance of the transition state to the folded state.  The value of
$\alpha$ is easily obtained from the data.  If we make the obvious
assumption that the dynamical factors are approximately the same for
the different sequences, then equation (\ref{linear_free_energy})
implies that a plot of the logarithm of the folding rate against the
logarithm of the equilibrium constant for folding will be a straight
line with a slope of $\alpha$~\cite{leffler}.  When we plot the
logarithm of the folding rate {\em versus} the logarithm of the
equilibrium constant for different sequences, we see such a nice
linear free energy relationship, shown in Figure~\ref{fig:marcus}.  At
the temperature $T=1.0$ the folding time seems nearly independent of
the driving force, while the driving force is entirely reflected in
the unfolding rate.  Thus folding here is nearly entirely
``downhill'', a Type 0 scenario.  (The large fluctuations suggest a
Type 0B.)  At $T=1.26$ there is a clear nucleation barrier, but it is
small.  The transfer coefficient of $\alpha = 0.1$ suggests a rather
early transition state, {\em i.e.}, at this temperature the bottleneck
configurations are collapsed but have little native structure.  The
further increase of $\alpha$ at higher $T$ reflects a later transition
state as the entropy terms become more important.  This shows that the
transitions are only weakly Type I and essentially Type 0 under these
thermodynamic conditions.  The success of this analysis is remarkable
because the native structures corresponding with the sequences are not
strongly related to each other unlike the situation in site-directed
mutagenesis experiments.

\begin{figure}
\insfig{1}{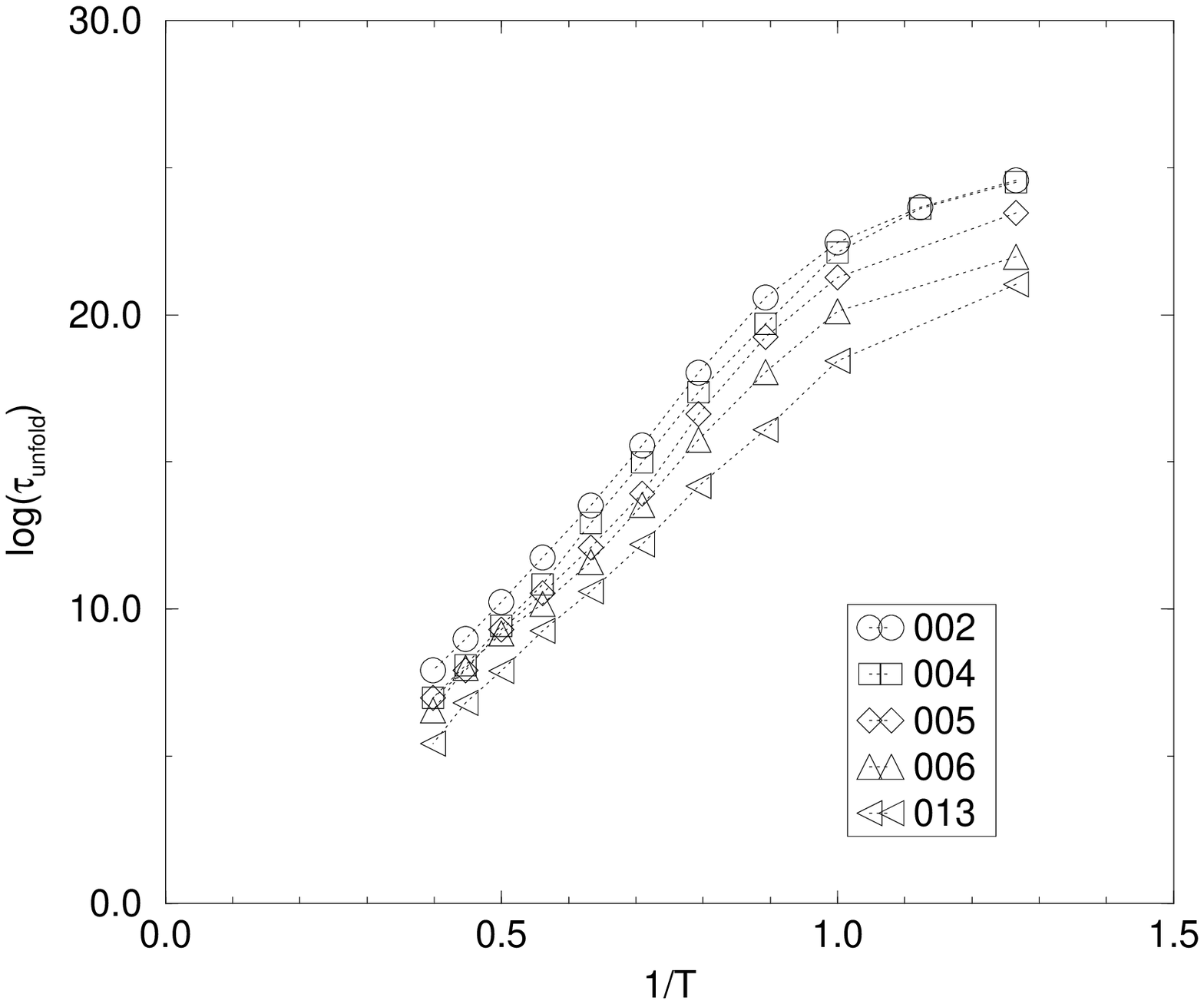}
\caption{
  An Arrhenius plot of the unfolding time against the inverse
  temperature for five of the sequences.  The unfolding time was
  calculated by multiplying the folding time by the ratio of the
  folded population to the unfolded population at a given temperature.
  Consequently, there is the same saturation effect at low temperature
  as in figure 9 caused by the finite simulation time.}
\label{fig:arrhen}
\end{figure}

An Arrhenius plot of the unfolding time {\em versus} $1/T$ is shown in
Figure~\ref{fig:arrhen}. This curve shows the dynamical effects as
$T_g$ is approached.  There is a clear change in the behavior of the
activation energy for unfolding near $T_g$ where the curve starts to
level off.  This behavior reflects the change in dynamics at $T_g$
suggested by the energy landscape analysis.  The loss of dynamical
flexibility caused by the entropy crisis leads to dynamical
reorganization times limited by the entropy of search and the
activation energy of the elementary step. (See equation
(\ref{search_time})) This analysis of the computer simulation shows
many of the ways in which data can be reduced when the thermodynamic
dependence of the underlying forces is understood.

\section{Energy Landscape Analysis and Folding Experiments}

We now turn to the analysis of some particular proteins that have been
studied extensively in the laboratory, lysozyme, chymotrypsin
inhibitor, and cytochrome {\it c}.  Despite the significant work
already done on these systems, we believe that there is insufficient
data to uniquely classify the mechanisms of folding via our energy
landscape framework.  However, it is possible to use the existing data
to give a flavor of how these ideas can be used in laboratory
situations.  As we have seen in our discussion of the computer
simulations, many qualitative features of experiments, such as curved
Arrhenius plots, can be obtained from the energy landscape scenario,
and can even be quantified if the underlying driving forces are
understood.  A considerable difficulty in the experimental studies is
that these driving forces are temperature dependent~\cite{leikin}.  It
is however important to realize that we can separately change the
driving force by such devices as the use of denaturant or mutation and
separate this effect from those effects which are directly due to the
ruggedness of the energy landscape due to thermal energies.
Ruggedness is a more nearly self-averaging quantity.  A further
analysis of this type for specific systems will, we hope, be made
soon.\footnote{At this point the reader may wish to review the folding
  scenarios discussed in Figure 5.}

In some ways, the simplest experimental situation occurs for those
proteins and conditions which exhibit a Type I folding mechanism.  The
kinetics in such systems should be simple exponential.  These systems
have moderate driving forces and are studied in the near equilibrium
range near the midpoint of the transition curve.  One feature favoring
a Type I transition as opposed to a Type II transition is the
avoidance of premature collapse.  When collapse occurs corresponding
changes in the ruggedness of the energy landscape can arise and play a
role.  Apparently Type I behavior occurs upon the cold denaturation of
lysozyme as studied by Chen and
Schellman~\cite{chen_schellman_1,chen_schellman_2}. It is
substantially a uniexponential process.

The folding of chymotrypsin Inhibitor 2, an 83 residue monomeric
protein with no disulfide bonds, has been studied by Jackson and
Fersht~\cite{jackson_fersht_1,jackson_fersht_2}.  In many respects
their experiments resemble a Type I scenario.  Jackson and Fersht used
fluorescence measurements and scanning microcalorimetry to study the
refolding of this protein.  The equilibrium denaturation experiments
found strong evidence for a simple two-state transition without
intermediates.  The kinetic measurements, however, reveal three
phases, but it is clear that these are due to the five proline
residues in the molecule, of which at least four are in the {\em
  trans} state in the crystal structure.  Seventy-seven percent of the
protein molecules fold with a time constant of $ .02$ seconds and the
two observable slow phases have time constants of $43$ and $500$
seconds.  The slow phases are catalyzed by peptidyl-prolyl isomerase,
which catalyzes proline isomerization.  The fast phase is not affected
by this enzyme.  The protein molecules that start with all the
prolines in the {\em trans} configuration have very nearly exponential
kinetics on the timescale studied.

In the energy landscape view, proline isomerization appears as a high
ridge separating the configurations with a {\em cis} isomer from those
with a {\em trans} isomer~\cite{nall_1,nall_2}.  One such ridge
appears for each proline in the protein.  Each of these separate parts
of the configuration space can be analyzed with the simple energy
landscape concepts that we have already discussed.  Thus, the mere
observation of multiexponentiality is not enough to imply that these
systems obey Type II kinetics in which a glass transition is present.
These ridges in the energy landscape come from the simple effect of
single amino acid residues, whereas the glass transition comes from
the composite effect of all the amino acid residues in the protein.

An example of apparent Type II behavior in provided by hen lysozyme at
its high temperature denaturation transition.  The evidence for Type
II behavior of lysozyme at this transition is largely based on the CD
measurements and pulsed hydrogen-exchange labeling carried out by
Radford {\em et al.}~\cite{radford}.  These studies suggest
multi-exponential behavior for the protection of the amide hydrogens,
which Radford {\em et al.} have interpreted as due to the existence of
multiple parallel folding pathways.  The Type II nature of this
transition apparently occurs because of the possibility of early
collapse.  In addition, misfolding is apparently present since the CD
shows, after the first $100$ milliseconds, considerably more
$\alpha$-helix present than is present in the native state.  Thus, in
this situation, the folding protein adopts a locally favorable
conformation which must be partly unfolded to get into the globally
favored native state.  The initial strong local tendency towards helix
formation is giving rise to frustration in the technical sense of
competing interactions discussed earlier in this paper.  The Type II
behavior suggests that the roughness of the energy landscape for
lysozyme is actually larger, compared to $k_B T$, at the high
temperatures than at the low temperatures, apparently due to the
temperature dependence of the hydrophobic forces.

Cytochrome c, with its heme constraints, apparently has little
roughness to its energy landscape compared to the free energy
gradient.  The heme is covalently bound to the protein chain and after
the iron coordination sphere is completed, folding of different parts
of the protein occur rather rapidly.  On the other hand, the heme
group can also be mis-ligated by some of the amino acids in the
protein and this misligation can be detected spectroscopically.  The
mis-ligated population can not follow the free energy gradient all the
way to the native structure so the presence of the heme also
facilitates the study of the different mis-folded structures present
in an ensemble of folding proteins.  Sosnick {\em et al.} have studied
the folding of cyctochrome {\em c} under conditions where the
misligation does not occur~\cite{tobin}.  They found that about 50 -
70 \% of the molecules in this population acquired native secondary
and tertiary structure with a time constant of approximately 15
milliseconds.  They estimated, from fluorescence quenching, the time
constant for collapse to be approximately 12 milliseconds, that is,
the of the same order as the folding time.  These experiments suggest
that cytochrome {\em c} folding is Type 0 under these conditions
though it is difficult to assign it to Type 0A or Type 0B with the
data from these experiments.  Experiments on cytochrome {\em c}
folding provide a good illustration of how the folding of a particular
protein can vary {\em qualitatively} as the conditions of the folding
experiment vary.  For example, when cytochrome {\em c} is refolded at
pH 6.2, the folding is multiexponential and takes on the order of
seconds.  Sosnick {\em et al.} have also shown that the slow folding
at pH 6.2 is due to the formation of a mis-folded, collapsed
structure, rather than the specific mis-ligation of the heme, in
agreement with the picture of a glassy phase presented here.

\section{Conclusion}

The energy landscape picture allows us to combine various disparate
ideas about the nature of biomolecular self-organization in protein
folding.  The energy landscape picture can accommodate multiple
parallel path scenarios, as well as unique, sequence-dependent
pathways for protein folding.  The crucial concept in understanding
particular experimental and computer simulation situations is to
organize the kinetics of the problem through the consideration of a
phase diagram and to study of the dynamics of the crucial order
parameters for folding which distinguish folded states from unfolded
ones.  In a generic energy landscape picture, several different phase
transitions occur and are coupled.  At the very minimum, one must
consider the two purely thermodynamic transitions of folding and of
collapse.  The collapse transition temperature depends upon both the
overall tendency for self-association and also on the ruggedness of
the energy landscape.  Above the glass transition collapse is a
largely self-averaging process; that is, it depends on the overall
composition of the sequence and on little else.  The folding
transition, on the other hand, is always sensitive to the details of
the sequence.  In addition to these conventional, understood phases, a
rough energy landscape exhibits a glass transition which occurs near a
thermodynamic glass transition temperature, $T_g$.  This temperature
is also a self-averaging property of different sequences of similar
composition.

Different scenarios for protein folding mechanisms occur, depending on
the relationship of these various temperatures and the conditions
under which the experiment is carried out.  The simplest situation to
understand occurs when there is a moderate driving force toward the
folded state.  Near to the midpoint of the denaturation curve, there
will be an overall double minimum potential of free energy function
and the roughness of the energy landscape simply acts to modulate the
rate of passing over the transition state.  This transition state is
actually a set of many configurations and could be said to consist of
numerous micro-transition states in a funnel toward the folded state.
The kinetics in this situation are simple exponential.  If the driving
forces for folding are considerably smaller, the folding temperature
can become close to the glass transition temperature.  In this case
one encounters considerable slowing of the folding process itself; a
Type II scenario emerges in which individual pathways for folding can
be dissected.  Here there will be multiple exponential processes
typically.  The great irony, of course, is that in the situation where
we can find individual pathways, folding will be typically very slow.
Indeed, nearly kinetically unfoldable proteins would exhibit the most
clearly defined pathway for folding.  These discrete pathways,
however, are not self-averaging aspects of the dynamics and are
sensitive to individual mutations in sequence.

For very large driving forces, one can encounter Type 0 scenario
folding in which essentially all of the dynamics goes on in a downhill
manner.  If a Type 0 scenario can occur much above $T_g$, this gives
rise to processes that are very fast (of order of ordinary homopolymer
collapse times)~\cite{degennes}.  On the other hand, if the glass
transition intervenes, which is likely if non-specific collapse
occurs, individual pathways can still be found, and, again, they will
be strongly sequence dependent and sensitive to mutations.

If the qualitative nature of the interaction energy scales is
understood, detailed temperature dependences can be obtained by the
energy landscape analysis.  A typical feature of this analysis is that
one obtains curved Arrhenius plots for folding times, much like those
actually occurring in experimental situations.  This curvature
reflects the roughness of the energy scale of the particular protein
and enters in both a thermodynamical and dynamical way.  The other
energy scale is related to the folding temperature itself and to the
stability gap in the energy spectrum of kinetically foldable proteins.
Simple linear free energy relations between the folding time and the
stability gap energy scale are obtained.  A most remarkable feature,
however, is that there are discontinuities in these relations and in
the apparent activated energies themselves as the glass transition is
approached.  The main difficulty in using energy landscape analysis to
interpret laboratory experiments is the temperature dependence of the
underlying thermodynamic forces.  Still, the self-averaging nature of
the roughness energy scales {\em versus} the specific sequence
dependence of the stability gap scale should allow some insight to be
obtained in real experiments.  The employment of different modes of
denaturation will be essential in differentiating these energy scales
of the protein folding landscape.  One can think of the use of
chemical denaturants, such as urea and guanidine, that largely bind to
unfolded configurations as primarily affecting the stability gap
rather than the roughness energy scale.  On the other hand, pressure
will strongly effect all solvent mediated forces and thus will
correlate with the roughness energy scale~\cite{zipp,li,sama,royer}.

Another complexity in laboratory experiments is that there con be
multiple order parameters for real proteins, since folded structures
differ in several ways from the typical unfolded ones.  The point is,
however, that there are probably only a few such parameters and a few
overall energy scales that are relevant.  If the dynamical
re-organization timescales for each of these order parameters are
similar the many reaction coordinate situation does not differ
dramatically from the one effective coordinate picture we have
discussed in detail in this paper.  If the timescale for different
motions differ appreciably, either through local energy barriers or
glass transition temperatures that vary with these order parameters, a
more complex scenario in which the folding bottleneck is largely
independent of the equilibrium free energy barrier can arise.  Still
the few coordinate generalization of the present analysis would be
applicable.  Experimentally this situation would resemble Type II or
Type 0B one coordinate scenarios, in that multi-exponential kinetics
would be prevalent.

The most important additional order parameters are those measuring the
degree of collapse, secondary structure, e.g. helical content, and
side-chain ordering.  The glass transition characteristics depend
greatly on collapse, so this is one possible source of decoupling of
the bottleneck from the equilibrium free energy
barrier~\cite{bw_biopolymers}. The ruggedness of the energy landscape
also can depend on side-chain orientation since some mis-associations
may simply not be sterically allowed for some sidechain orientations.
In addition, the configurational entropy of the backbone depends on
its helical content, again affecting the dynamical glass transition.
Certainly in multi-domain proteins one must use different reaction
coordinates for each folding unit.  Even single domain proteins may
have different folding substructures.  Some analyses such as that of
Bryngelson and Wolynes, suggest that the critical nucleus for folding
is large ~\cite{bw_biopolymers} but other studies suggest smaller
sizes for the critical nucleus and concomitantly smaller folding units
with separate reaction coordinates~\cite{dave_1}.  In any case, an
energy landscape analysis allows us to reduce, in many circumstances,
a huge number of variables down to only a few degrees of freedom and a
statistical characterization of the roughness of the energy landscape.
The true diversity of the energy landscape only comes through in the
Type II scenarios in which the glass transition has intervened.  A
study of most experiments suggests that many proteins are near to the
glass transition and may show Type 0B and Type II scenarios.  Since
the roughness of the energy scale is self-averaging, it will be
interesting to explore the phase diagram for different protein and
especially to examine different protein compositional classes to see
if there are systematic differences in energy scale roughness in {\em
  in vitro} folding.

One of the major fruits of the energy landscape analysis of protein
folding has been a simple variational criterion for achieving
fast-folding proteins.  The minimal frustration principle, which at
first seemed a qualitative concept, has been formulated now as a
criterion for the maximization of the folding temperature compared to
the glass transition temperature.  This principle has already been
used to reverse engineer proteins to discover correlations that are
important in predicting protein structure~\cite{glsw1,glsw2}.  In
addition, it has been used to design proteins that can fold on
reasonable timescales on computers~\cite{sg_pnas}.  It will be
interesting to see whether the combination of the reverse engineering
and engineering approaches will allow the design of kinetically
foldable proteins in the laboratory.

\section{Acknowledgments}

We gratefully acknowledge useful discussions with and encouragement
from William A. Eaton, Hans Frauenfelder, V. Adrian Parsegian and
Attila Szabo.  The useful comments on an earlier version of the
manuscript by Hue Sun Chan, Ken A. Dill and Donald M. Engelman are
also appreciated.  J.~D.~B. thanks the Division of Computer Research
and Technology of the National Institutes of Health for supporting his
research.  J.~N.~O. is a Beckman Young Investigator. The work in San
Diego has been funded by the Arnold and Mabel Beckman Foundation and
by the National Science Foundation (Grant No.\ MCB-93-16186).
J.~N.~O. is in residence at the Instituto de F\'{\i}sica e
Qu\'{\i}mica de S\~ao Carlos, Universidade de S\~ao Paulo, S\~ao
Carlos, SP, Brazil during part of the summers.  P.~G.~W's research is
supported by NIH (Grant No.\ 1 R01 GM44557).  This work was started at
the Institute for Theoretical Physics, University of California, Santa
Barbara, during a mini-program on Protein and Nucleic Acid Folding in
January 1993, supported by the NSF (Grant No.\ PHY-89-04035).

\end{document}